\documentclass[prb,preprint,amsmath,amssymb]{revtex4}
\begin{document}
\author{Kevin Leung,$^{1*}$ Fernando Soto,$^2$ Kie Hankins,$^2$
Perla B.~Balbuena,$^2$ and Katharine L.~Harrison$^1$}
\affiliation{$^1$Sandia National Laboratories, MS 1415, Albuquerque, NM 87185\\
$^2$Department of Chemical Engineering, Texas A\&M University, College
Station, TX 77843\\
$^*${\tt kleung@sandia.gov}}
\date{\today}
\title{Stability of Solid Electrolyte Interphase Components on Lithium Metal
and Reactive Anode Material Surfaces}

\input epsf

\begin{abstract}

Lithium ion batteries (LIB) can feature reactive anodes that operate at low
potentials, such as lithium metal or silicon, passivated by solid electrolyte
interphase (SEI) films.  SEI is known to evolve over time as cycling proceeds.
In this modeling work, we focus on the stability of two main SEI components,
lithium carbonate (Li$_2$CO$_3$) and lithium ethylene dicarbonate (LEDC). 
Both components are electrochemically stable but thermodynamically unstable
near the equilibrium Li$^+$/Li(s) potential.  Interfacial reactions represent
one way to trigger the intrinsic thermodynamic instability.  Both 
Li$_2$CO$_3$ and LEDC are predicted to exhibit exothermic reactions on lithium
metal surfaces, and the barriers are sufficiently low to permit reactions on
battery operation time scales.  LEDC also readily decomposes on high
Li-content Li$_x$Si surfaces.  Our studies suggest that the innermost SEI
layer on lithium metal surfaces should be a thin layer of Li$_2$O -- the
only thermodynamically and kinetically stable component (in the absence
of a fluoride source).  This work should also be relevant to inadvertant
lithium plating during battery cycling, and SEI evolution on Li$_x$Si surfaces.

\end{abstract}

\maketitle

\section{Introduction}

Solid-electrolyte interphase (SEI) films that passivate low voltage
anode surfaces are critical for lithium ion battery
operations.\cite{book2,book1,book,review,novak_review,intro1}  These films
arise from electrochemical reduction and subsequent breakdown of the organic
solvent-based electrolyte and additive molecules, which are unstable under
battery charging potentials.  The SEI blocks further electron transfer from
the anode to the electrolyte, yet permits lithium ion (Li$^+$) transport.  
SEI films on traditional graphite anodes are not static, but grow thicker
during cycling.  Moreover, they are known to evolve either by dissolution or
chemical processes.\cite{evolv1,evolv2,evolv3,evolv4}  Silicon, a potential
next-generation anode material,\cite{si_review} can expand by 300\% upon
full lithiation, and the SEI films coating its
surfaces\cite{chan1,chan2,kim2007,aurbach,lucht1,lucht2,edstrom1,edstrom2,edstrom3} are likely to crack, delaminate,\cite{fracture,sullivan} and in general
exhibit more dynamic evolution than on graphite surfaces.  
Lithium metal,\cite{aurbach93,aurbach99,li_review}
much more reactive than graphite, is also being considered as transportation
battery anode material.  Hence there is an urgent need for further
understanding of SEI evolution and stability on these reactive anode surfaces.

In terms of SEI stability, SEI dissolution in liquid electrolytes has
been discussed,\cite{dissolves} as has the chemical instability of SEI
components due to elevated temperature,\cite{thermal1,thermal2} the presence
of acid,\cite{acid} and reactions with transition metal incorporated into
the SEI.\cite{nidiffus,mndiffus}  A recent pioneering computational work that
emphasizes the concept of SEI stability has predicted that organic SEI
components can react with Li$_x$Si surfaces, as well as undergo attacks by
radicals present in the electrolyte.\cite{soto}  These interfacial reactions
between SEI components and active materials help determine the chemistry at
electrode/SEI interfaces. Understanding such reactions is crucial for 
generating detailed models of SEI structures that govern their
electron-blocking and Li$^+$ transmitting functions.

The present work systematically addresses the stability of multilayer organic
and inorganic SEI components on Li metal and Li$_x$Si surfaces.  This work is
motivated by solid-solid interface modeling conducted in the context of
lithium ion battery solid electrolytes,\cite{sodeyama,holzwarth1,holzwarth2,santosh,sumita,ong,mo,mo1}
and assumes a time lapse between the first deposition of SEI products and
their destruction.  Thus it is complementary to the liquid-solid interface
approach\cite{soto} more pertinent to the initial stages of SEI growth.

First we distinguish three possible criteria of SEI stability: thermodynamic,
electrochemical, and interfacial.  (1) Following Ref.~\onlinecite{phasediagram},
we define ``thermodynamic instability'' to mean that the bulk SEI material
(assuming it can be synthesized in bulk quantity) can be decomposed into
one or more solid or gas phases with lower overall free energy.  This
criterion allows breaking any chemical bond and is agnostic about reaction
kinetics.  (2) ``Electrochemical instability'' is taken to mean that the
phase can be readily electrochemically oxidized or reduced during cyclic
voltametry.\cite{redox}  Only the cleavage of covalent or ionic bonds that
can break in the time scale of electron transfer is permitted, although
further work is needed to refine the precise meaning of the kinetic constraint.
Electrochemical stability is related to band-alignment diagrams frequently
invoked in the battery literature,\cite{phasediagram,goodenough} where 
redox reaction are assumed to begin when the Fermi levels in the metallic
electrode coincide with the valence or conduction band edges of the
electrolyte.  However, organic electrolytes, SEI components such as
Li$_2$CO$_3$, and many cathode oxide materials form localized charge states
(``states in the gap''), or small polarons, upon being reduced or oxidized.
Thus redox potentials, not band edge positions, are the rigorous governing
quantities.  Note that some authors describe this instability criterion as
``thermodynamic''\cite{shaohorn} rather than ``electrochemical'' (this work,
Ref.~\onlinecite{phasediagram}, and others).  (3) ``Interfacial
instability'' here means that exothermic reactions can occur at the interface
within typical lithium ion battery charge/discharge time scales.  When a SEI
component is thermodynamically unstable but electrochemically stable -- 
thus preventing long-range electron transfer -- interfacial reactions
represent one key way to trigger the intrinsic thermodynamic instability.  In
this work, we impose a one-hour reaction time threshold; short reaction times
are considered to yield interfacially unstability.

This work focuses on two dominant SEI components, namely inorganic lithium
carbonate (Li$_2$CO$_3$) and organic lithium ethylene dicarbonate (LEDC).  Both
components are widely reported to be significant components of the anode SEI
when the liquid electrolyte contains ethylene carbonate (EC) and other linear
carbonates.\cite{book1,book,review} Crystalline Li(100) and amorphous silicon
(a-Si) surfaces are used to represent two extremes of reactive anode
surfaces on which the SEI may decompose.  We stress that the a-Si/SEI
interface is intended to represent SEI-covered Si anode surface after cycling,
not pristine Si surfaces which should be covered with native
oxides.\cite{sio2_bal} The SEI/Li(100) interface is relevant to undersirable
Li-plating on graphite anodes\cite{plating1} and intentional Li-deposition
when Li metal is used as anode.\cite{plating2}  Li(100) may also be a
reasonable proxy for high $x$ Li$_x$Si surfaces because in previous modeling
studies,\cite{fec} the low energy surfaces of high-$x$ Li$_x$Si are found
to be dominated by Li atoms.  Si outcroppings on Li(100) are considered for
its effect on SEI reactivity in the supporting information document (S.I.).

It has generally been accepted that anode SEI consists of an inner inorganic
layer (e.g., Li$_2$CO$_3$, Li$_2$O, and/or LiF) and an outer organic layer
(e.g., LEDC and polymeric species).\cite{review,aurbach99}  However, the
LEDC/Li(100) interface is relevant for reasons not just related to inadvertant
Li-plating on the outer SEI surface.\cite{plating1}  A recent ultra-high vacuum
(UHV) study has shown that LEDC can form from a sub-monolayer of EC molecules
on Li metal surfaces at low temperature.\cite{robey} This finding is in
agreement with a modeling work suggesting that LEDC may be formed by
two-electron reduction of EC during the initial stages of SEI
formation,\cite{e2} not (only) at late stages via one-electron reduction.
Hence the reactions of LEDC on active material surfaces, presumably to 
produce inorganic species, are also relevant to the anode/SEI interface.

The voltage dependences of these processes will be discussed.  By ``voltage''
we refer to the electronic potential,\cite{solid} which depends on
the Fermi level and reflects the instantaneous applied potential.  This
definition is consistent with computational studies in many electrochemical
applications.\cite{sprik12,gross08,neurock06,rossmeisl15,galli,otani12,arias12,hybertsen} In contrast, the lithium chemical potential, which is a function of
the Li-content, is a slow-responding property that indicates whether the
interfacial structure is out-of-equilibrium or not; at interfaces it
does not by itself determine the instantaneous voltage.\cite{solid}

{\it Ab initio} molecular dynamics (AIMD) simulations are also conducted to
show that decomposition reactions of organic SEI components similar to
EDC can readily occur in less well-controlled, liquid-solid
interface environments, with multiple species decorating the surfaces of
Li$_x$Si crystalline anode with finite Li-content -- instead
of just Li(100) or Si-doped Li-metal surfaces.  These
simulations examine the decomposition of lithium vinylene dicarbonate
(Li$_2$VDC) during the initial stages of SEI formation, when liquid
electrolyte, LiF clusters, and submonolayer Li$_2$VDC are all present.  VDC
differs from EDC only by a carbon-carbon double bond and two less protons.
It can be formed in the presence of vinylene carbonate, a well-known
electrolyte additive.  The LiF cluster represents another surface ``defect''
on otherwise crystalline Li$_x$Si.  The AIMD predictions in these systems
complement and broadly agree with our static, solid-interface results.
The reactions of Li$_x$Si with the other SEI component considered herein,
Li$_2$CO$_3$, are more challenging due to the need for lattice matching between
crystalline Li$_2$CO$_3$ and the large Li$_x$Si surface unit
cells.\cite{chev,greeley,hwangut}  Such studies will be conducted in the future.

This paper is organized as follows.  Section~\ref{method} discusses the details
of the DFT calculations.  Section~\ref{results} describes the results.  The
implications of the predictions and comparison with experiments are discussed
in Sec.~\ref{discussions}, and Sec.~\ref{conclusions} summarizes the paper.

\section{Methods}
\label{method}

We apply periodic boundary condition (PBC) DFT methods to assess
thermodynamic stability and interfacial stability, and cluster-based
(CB) DFT calculations to examine electrochemical stability.  An SEI
component is defined to be thermodynamically unstable if energy is
gained by transforming it into other phases upon reaction with Li metal;
to exhibit interfacial instability on a particular anode material
surface if it reacts exothermically within one hour; and to exhibit
electrochemical instability if CB-DFT predicts a reduction potential
which is positive relative to the Li$^+$/Li(s) reference.  If an SEI
component reacts on an anode surface and transforms into a new stable
phase or film which passivates the anode, the original SEI component is
still regarded as unstable at the interface.

\begin{figure}
\centerline{\hbox{ \epsfxsize=1.4in \epsfbox{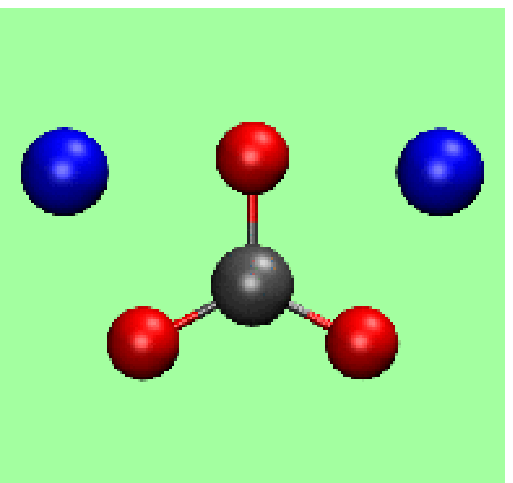}
                   \epsfxsize=1.4in \epsfbox{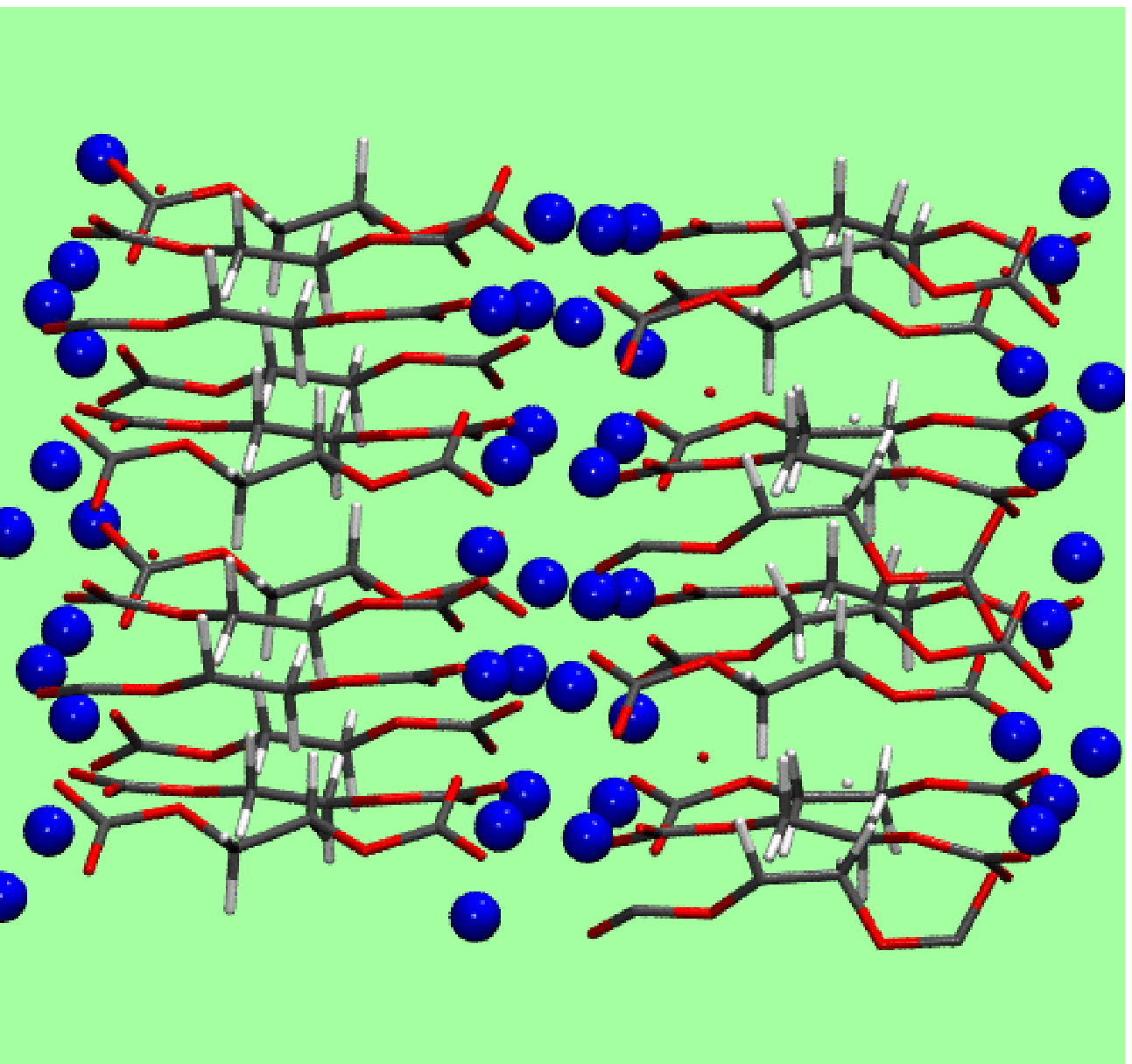} }}
\centerline{\hbox{ (a) \hspace*{1.19in} (b) }}
\centerline{\hbox{ \epsfxsize=1.4in \epsfbox{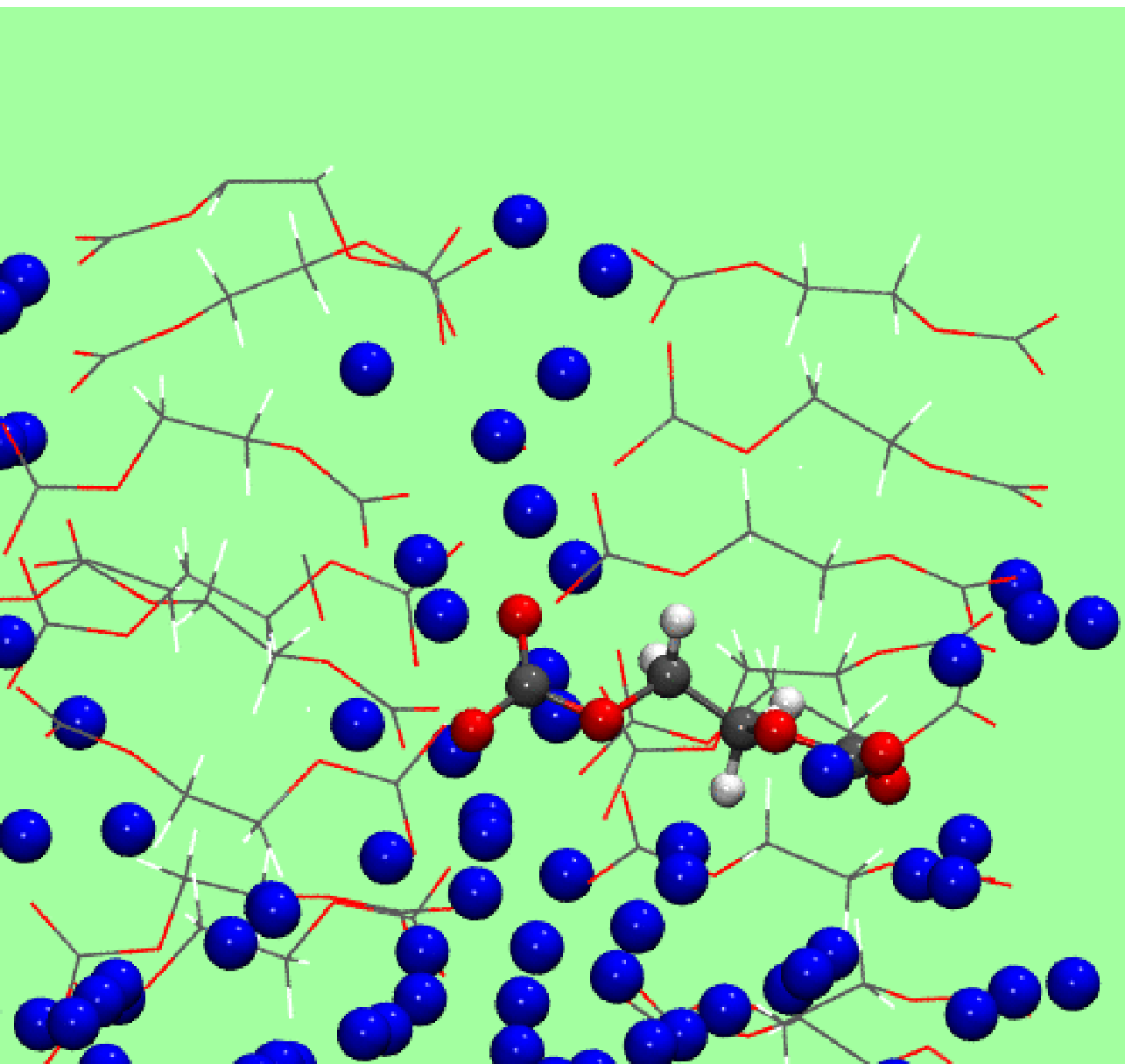}
                   \epsfxsize=1.4in \epsfbox{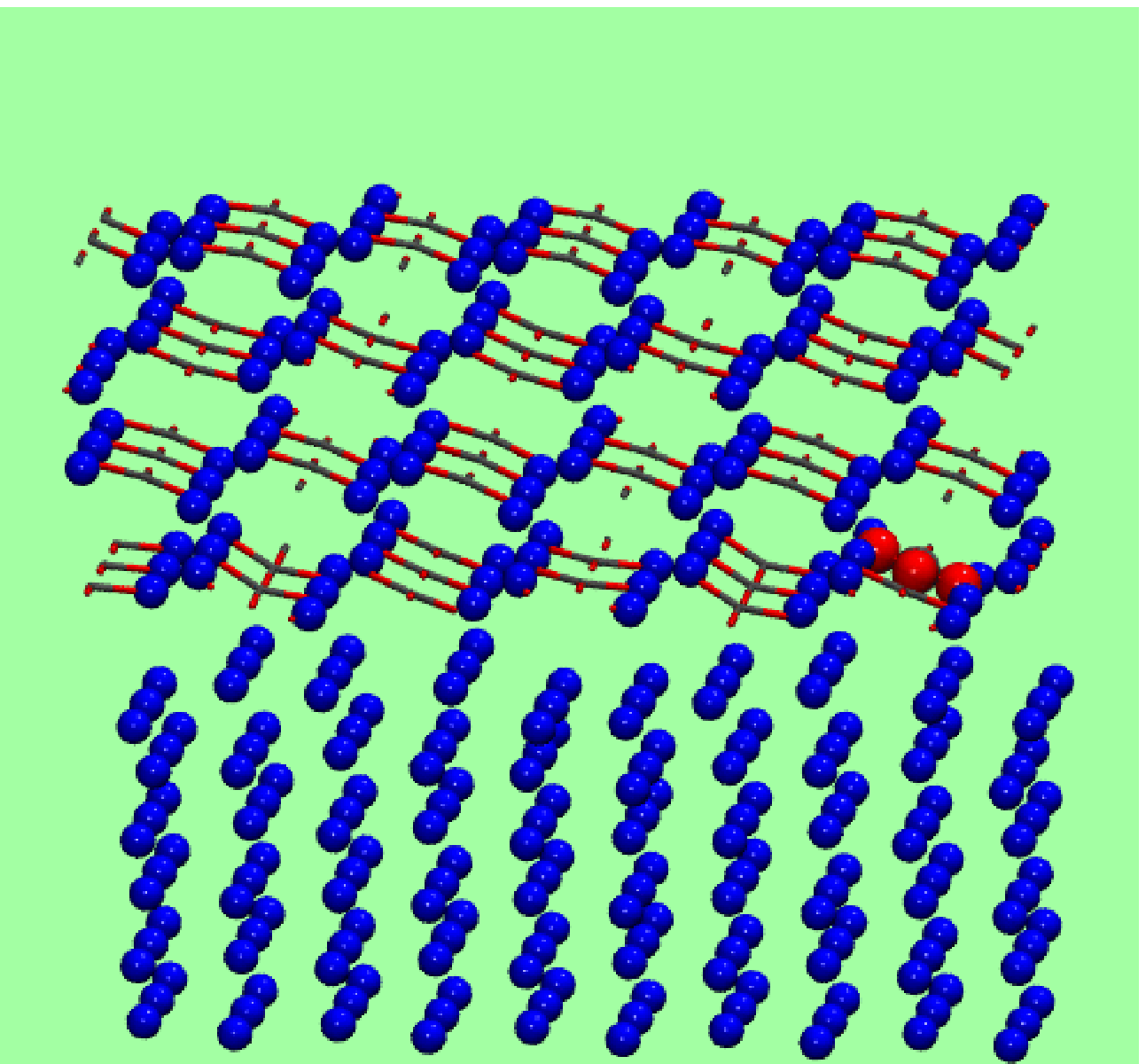} }}
\centerline{\hbox{ (c) \hspace*{1.19in} (d) }}
\centerline{\hbox{ \epsfxsize=3.2in \epsfbox{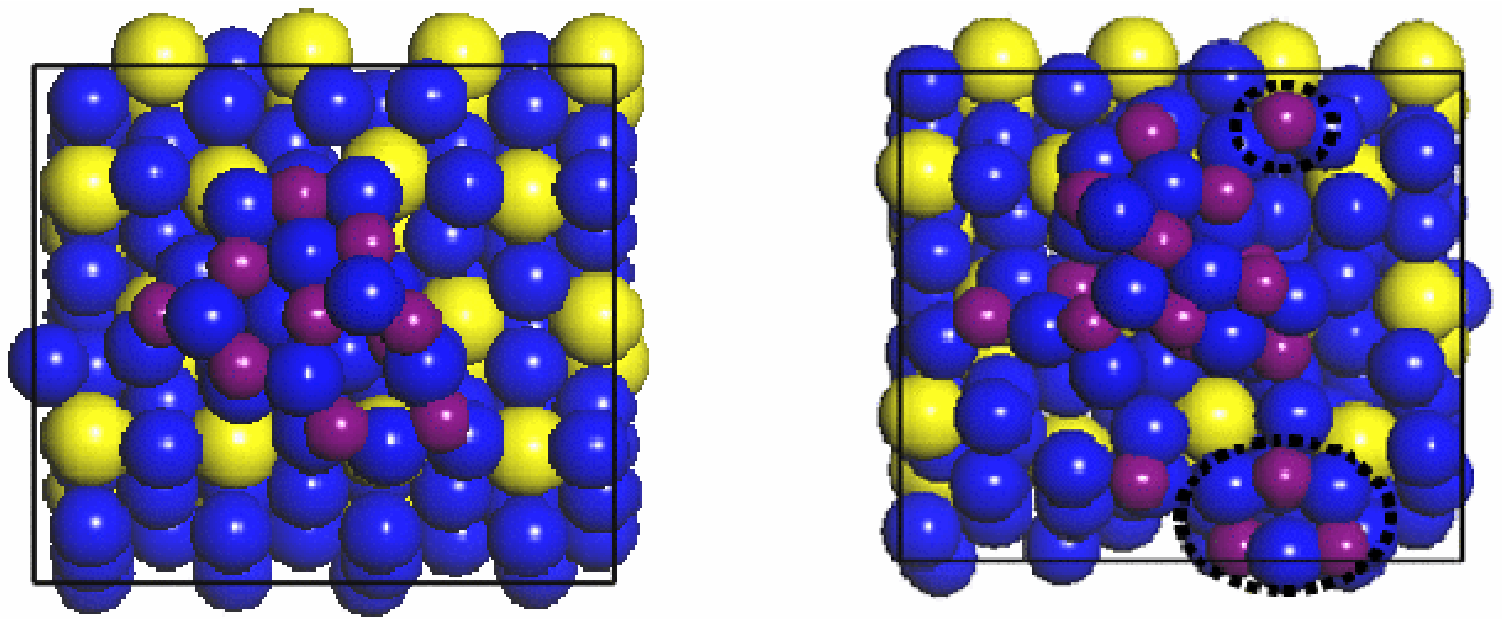} }}
\centerline{\hbox{ (e) \hspace*{1.19in} (f) }}
\caption[]
{\label{fig1} \noindent
(a) Li$_2$CO$_3$ formula unit. (b) Li$_2$C$_4$H$_4$O$_6$ (LEDC) ``crystal,''
optimized using DFT/PBE calculations.  (c) 3-bilayer LEDC on Li(100).  The
strong interaction renders the initially crystalline Li surface almost
amorphous.  (d) 4-layer Li$_2$CO$_3$(001) on Li(100).  Blue, grey, red, and
white represent Li, C, O, and H atoms, respectively.  Li, and molecular species
which react with the anode surface, are depicted as ball-and-stick figures;
non-reacting species are usually sticks or lines.  (e) Top view of LiF cluster
adsorbed on Li$_{13}$Si$_4$ (010) surface.  (f) The smaller sets of LiF~pairs
are shown inside the black circles.  They result from PF$_6^-$ decomposition or
FEC reduction after 9.6~ps AIMD simulations (Sec.~\ref{aimd}).  Yellow spheres
depict Si atoms in panels (g)\&(h).
}
\end{figure}

PBC DFT calculations are conducted using the Vienna Atomic Simulation Package
(VASP) version 5.3.\cite{vasp,vasp1,vasp2}  Most calculations apply the
PBE functional.\cite{pbe}  HSE06,\cite{hse06} which exhibits less 
delocalization error,\cite{wtyang} is used to re-examine systems with marginal
kinetic stability.  A 400~eV planewave energy cutoff is applied in all cases.
Since previous AIMD simulations of the initial stages of SEI formation have
rarely yielded CO$_3^{2-}$ and have predicted no LEDC,\cite{fec,martinez}
configurations from AIMD trajectories are not used as input.  Instead, we have
constructed multi-layer Li$_2$CO$_3$ (001) and LEDC slabs on anode
material surfaces.  Several simulation cells are depicted in Fig.~\ref{fig1},
with further details listed in Table~\ref{table1}.  The climbing-image
nudged elastic band method (NEB) is applied to predict reaction
barriers.\cite{neb}  A predicted reaction time scale of 1-hour requires that
the total net reaction exhibits a negative free energy change and sufficiently
low reaction barrier(s).  If a typical vibrational frequency-related
prefactor of $k_o$=10$^{12}$/s is assumed, the overall rate
$k = k_o \exp(-\Delta G^*/k_{\rm B}T)$ is faster than 1/hour when all reaction
barriers ($\Delta G^*$) are at most 0.92~eV.

\begin{table}
\begin{tabular}{||l|r|r|r|c||} \hline
system & dimensions & stoichiometry & {\it k}-point & figure \\ \hline
EDC/Li(100) 		& 19.50$\times$19.50$\times$34 & 
	Li$_{228}$C$_{72}$O$_{108}$H$_{72}$ & 2$\times$2$\times$1 & 1d,3,6,7\\
1-CO$_2$/Li(100)	& 30$\times$9.75 $\times$9.75 & 
	Li$_{48}$C$_{1}$O$_{2}$ & 1$\times$2$\times$2 & 5 \\ 
EDC/a-Si		& 16.75$\times$15.09$\times$38 &  
	Si$_{127}$Li$_{38}$C$_{72}$O$_{108}$H$_{72}$ & 1$\times$1$\times$1 & 7\\
Li$_2$CO$_3$(001)/Li(100)	& 25.02$\times$15.03$\times$34 &
	Li$_{324}$C$_{72}$O$_{216}$ & 1$\times$2$\times$1$^*$ & 1e,8a \\
1-Li$_2$CO$_3$/Li(100)	& 30$\times$9.75 $\times$9.75 & 
	Li$_{52}$C$_{1}$O$_{3}$ & 1$\times$2$\times$2 & 8e,8f \\
Li$_2$CO$_3$(001)/a-Si	& 16.68$\times$15.03 $\times$34 & 
	Si$_{127}$Li$_{96}$C$_{48}$O$_{144}$ & 2$\times$2$\times$1 & 1g,8a\\ 
Li$_2$VDC/Li$_{3.25}$Si(010)& 17.80$\times$15.94 $\times$34 & 
    Li$_{127}$Si$_{32}$F$_{48}$C$_{89}$O$_{99}$H$_{79}$ & 2$\times$2$\times$1 &
		1h,9\\ 
Li$_2$VDC/Li$_{3.25}$Si(010)& 17.80$\times$15.94 $\times$34 & 
    Li$_{127}$Si$_{32}$C$_{20}$O$_{30}$H$_{10}$ & 2$\times$2$\times$1 & 10\\ 
\hline
\end{tabular}
\caption[]
{\label{table1} \noindent
Simulation cell size (\AA$^3$), stoichimetry, and Brillouin zone sampling of
some periodic boundary condition model systems considered in this work.
``1-X'' means only one formula unit of ``X'' is present.  $^*$Increasing the
$k$-point sampling to 2$\times$3$\times$1$^*$ changes the CO$_3^{2-}$
decomposition barrier by less than 0.03~eV.
}
\end{table}

The standard dipole correction is applied to negate image interactions in
the periodically replicated, charge-neutral simulation cells.\cite{dipole_corr} 
For a simulation cell with a metallic Li(100) slab and a vacuum region,
$E_{\rm F}$ is well defined; the work function ($\Phi$) is the difference
between $E_{\rm F}$ and the vacuum level, and the electronic voltage is simply
($\Phi/|e|$-1.37)~V vs.~Li$^+$/Li(s), where $|e|$ is the electronic charge.
The ``1.37'' V value reflects the potential difference between an electron
at infinity and inside a Li metal foil held at 0~V; it is independent of the
SEI component.  This electronic voltage is to be directly compared to the
ideal applied experimental potential, i.e., it is assumed no ohmic loss exists.

Amorphous Si (a-Si) slabs are created by melting LiSi crystals at high
temperature, quenching, cutting a surface, and gradually removing Li.
a-Si slabs are insulating.  The relevant $E_{\rm F}$ in this case should
depend on impurity states inside the band gap.  Since this work does not
focus on defects, the voltage of a-Si cannot be unambiguously defined.  Future
modeling effort can address a-Si electronic voltages by doping with P and B
atoms to create metallic a-Si models.\cite{chan_dope}  The effect of
varying the charging rate can be qualitatively examined by setting unequal
ionic and electronic voltages, thus imposing an overpotential on the system.

The VASP code is also used to conduct AIMD simulations with Li$_2$VDC on
anode surfaces.  These simulations are conducted at T=450~K, using the NVT
ensemble and a 1~fs time step.\cite{martinez}  The simulation cell contains
a (010) Li$_{13}$Si$_4$ slab
and has dimensions 17.80$\times$15.94$\times$25.00~\AA$^3$.  Note that the
initial packing of the FEC molecules onto the anode surface containing an
overlayer of LiVDC is created using the BIOVIA Materials Studio Amorphous Cell
module.\cite{studio}  The LiF unit cell from the crystallography open
database\cite{database} and the crystal builder option of the Materials
Studio software are used to build a LiF model cluster on top of the anode
surface (Fig.~\ref{fig1}e-f).  A LiF cluster with a 5~\AA\, radius, consisting
of a total of 18~Li and 18~F atoms, is selected for this study.  The size
was chosen to represent a seed of the nucleating LiF phase; other cluster sizes
will be considered in the future.   A~1.0~M LiPF$_6$ salt immersed in a pure
FEC solvent (28~FEC molecules and two LiPF$_6$) is used as the electrolyte
solution that fills the gap between the two Li$_{13}$Si$_4$ surfaces.  As a
first step, we employ the classical universal force field (UFF)\cite{rappe}
and the Forcite module in BIOVIA to pre-equilibrate the electrolyte portion of
the system.  The electronic voltage is not calibrated but is expected to be low.

Cluster-based calculations are conducted using the Gaussian (G09) suite of
programs,\cite{g09} the DFT/PBE0 function,\cite{pbe0} and the ``SMD''
dielectric continuum approximation.\cite{smd}  Liquid electrolytes with
varying amounts of EC and dimethyl carbonate (DMC) or diethyl carbonate (DEC)
used in batteries exhibit slightly different $\epsilon_o$.  In this work, we
have followed Ref.~\onlinecite{e2} and used a single $\epsilon_o$=40 value.
Geometry optimization is performed using a {\tt 6-31+G(d,p)} basis set.
Single point energies are computed at a {\tt 6-311++G(3df,2pd)} level of
theory.  Vibrational frequencies are computed using the smaller basis,
yielding zero point energies and thermal corrections.

\section{Results}
\label{results}

\subsection{Thermodynamic Instability}

First we consider the following solid or solid/gas phase reactions at zero
temperature:
\begin{eqnarray}
{\rm Li}_2{\rm CO}_3 + 4 {\rm Li} \rightarrow 3 {\rm Li}_2{\rm O} + {\rm C} ; 
						\label{eq1}  \\
{\rm Li}_2{\rm C}_4{\rm O}_6{\rm H}_4  + 10 {\rm Li} 
\rightarrow 6 {\rm Li}_2{\rm O} + 4 {\rm C} +3 {\rm H}_2 ({\rm g}). \label{eq2} 
\end{eqnarray}
Here Li$_2$C$_4$O$_6$H$_4$ is the chemical formula for a LEDC unit.
All species are solids except those labeled as gas (``(g)'').  Diamond is used
to represent solid carbon.  Its cohesive energy is similar to the slightly
more stable graphite.  The DFT method used herein is not dispersion-corrected
and can underestimate graphite stability.  We stress that it is unnecessary
to find the most stable products; the existence of one set of product phases
lower in free energy than the original renders the latter thermodynamically
unstable.  The solid Li$_2$CO$_3$ structure is well known,\cite{li2co3} as
is body-centered-cubic Li.  There are 2.6~\% and 2.8~\% lattice mismatches
between the 5$\times$3 Li(001) and 3$\times$3 Li$_2$CO$_3$(001) slabs in the
lateral directions (Table~\ref{table1}), and lithium metal
is stretched by these amounts.  No crystal structure is available for LEDC,
and lattice matching information cannot be inferred.  However, molecular
dynamics simulations have revealed a layered structure.\cite{borodin12}  Our
3-bilayer LEDC simulation cell (Fig.~\ref{fig1}c) is obtained by optimizing
two staggered layers of LEDC molecules.

From DFT/PBE calculations, Eqs.~\ref{eq1}~and~\ref{eq2} are exothermic by
1.29~eV and 1.22~eV per Li consumed.  The fundamental reason for 
Li$_2$CO$_3$ instability is the fact that carbon atoms with formal charge
states of (+4) are in the presence of the extremely electronegative lithium
metal.  Finite temperature effects, ignored herein, may make Eq.~\ref{eq2}
even more favorable due to gas phase entropy production.  Replacing ``C'' with
Li$_2$C$_2$ also changes the energetics.\cite{lqchen} Extrapolating from these
predictions, it is likely that many carbon-containing SEI components with
high formal charges on carbon are thermodynamically unstable 
in the presence of low-potential, lithium-containing anodes.

In the formulation of Ref.~\onlinecite{phasediagram}, Eqs.~\ref{eq1}
and~\ref{eq2} should correspond to equilibrium reduction potentials of 1.29
and 1.22~V vs.~Li$^+$/Li(s) for Li$_2$CO$_3$ and LEDC.  However, the breaking
of many ionic or covalent bonds are needed to achieve the phase transitions,
which may consequently be kinetically hindered.  To our knowledge, no reduction
signature for Li$_2$CO$_3$ and LEDC at $>1$~V has been reported in
cyclic voltametry.  We note that CV curves can be difficult to analyze,
as they convolve contributions from electrochemical reactions of
fluoride-containing binders and metal oxides.  However, Li$_2$CO$_3$ obtained
from CoCO$_3$ can be electrochemically reduced to Li$_2$O and carbide
via conversion reactions.\cite{lqchen}  

\subsection{Electrochemical Stability}

Another definition of instability for Li$_2$CO$_3$ and LEDC is related to
their reduction potential ($\Phi$). $-|e|\Phi$ gives the (free) energy gained
when injecting an electron from an electrode held at 0~V vs.~Li$^+$/Li(s).

\begin{figure}
\centerline{\hbox{ \epsfxsize=1.5in \epsfbox{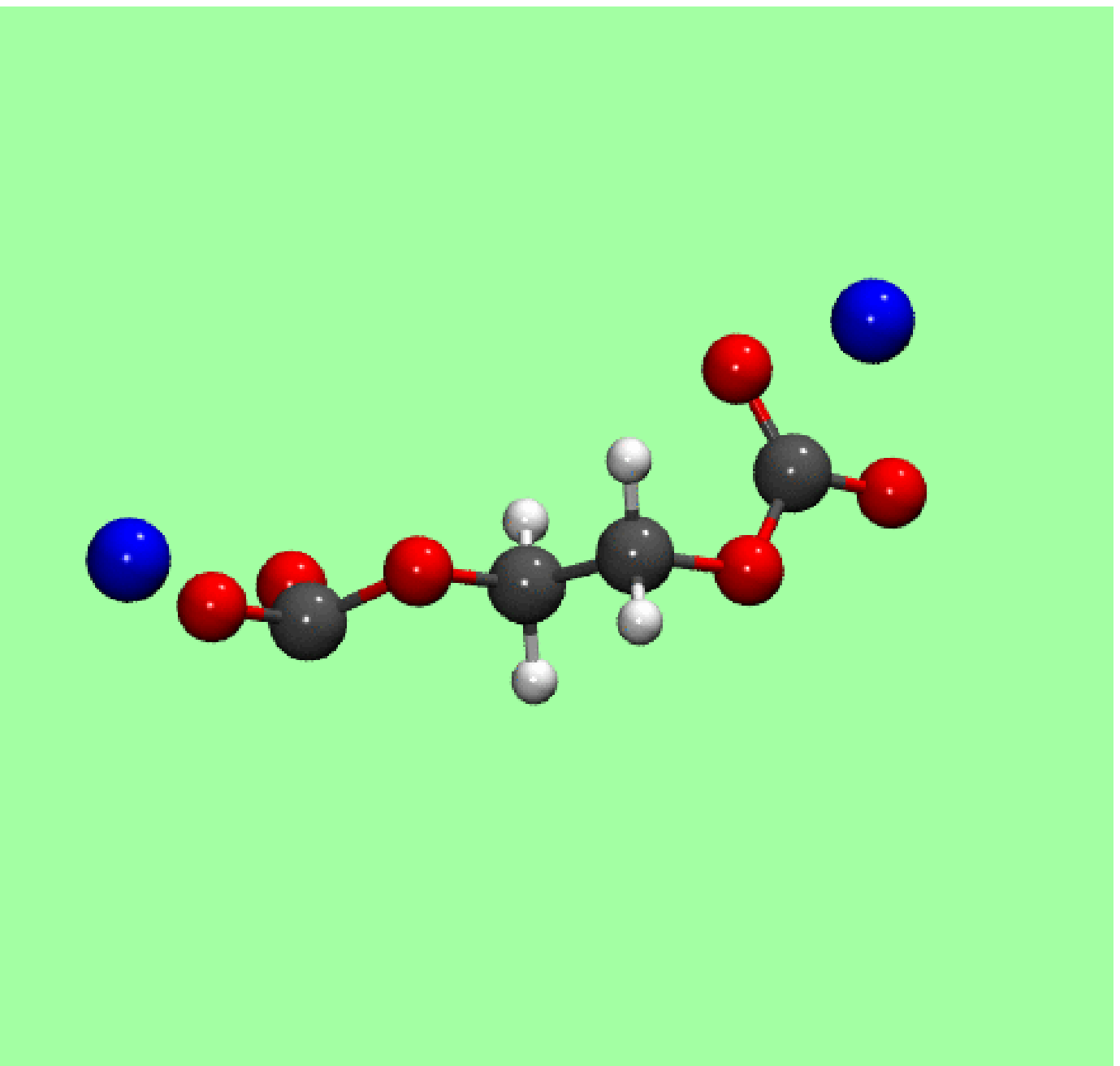}
                   \epsfxsize=1.5in \epsfbox{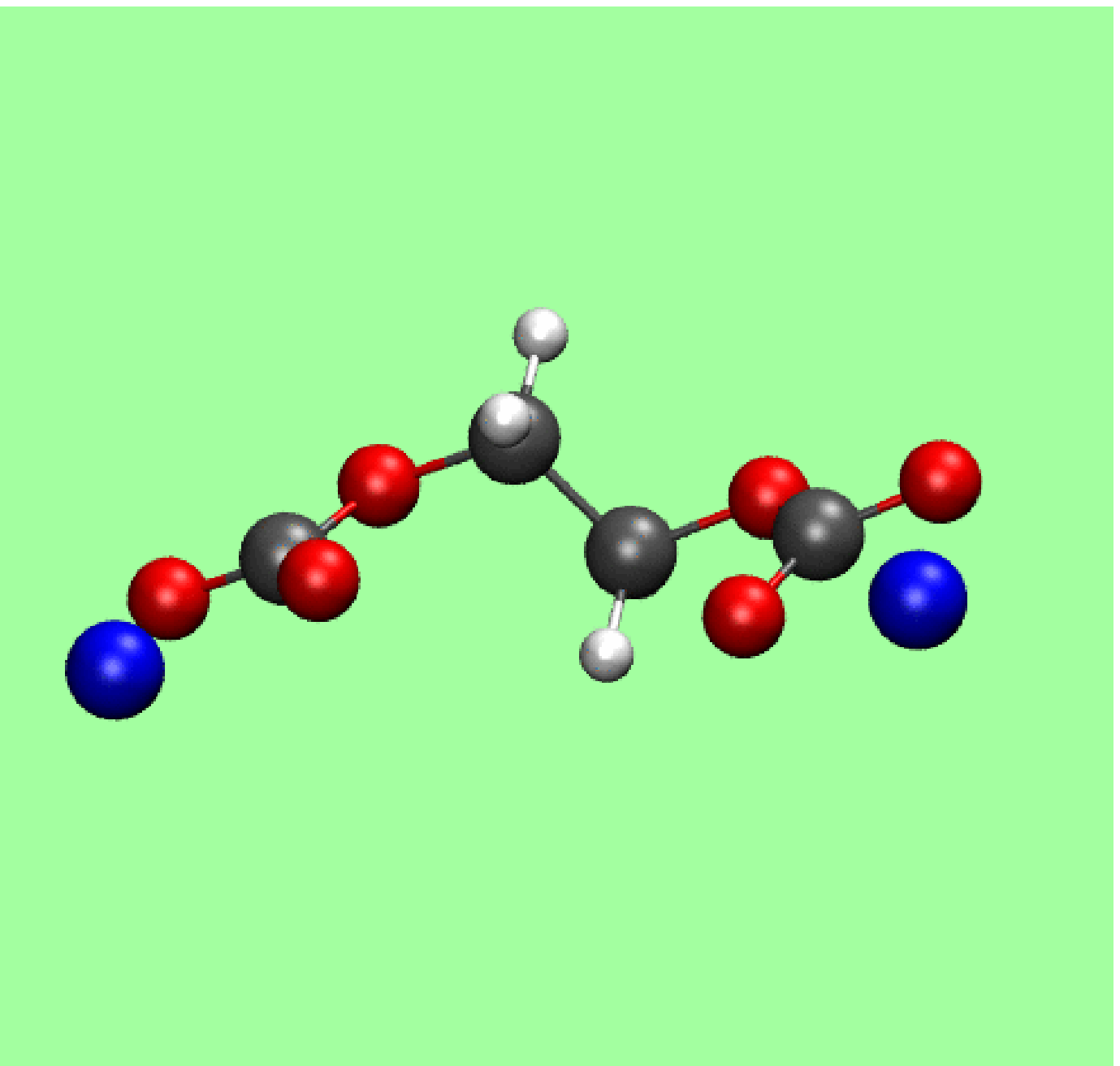} }}
\centerline{\hbox{ (a) \hspace*{1.29in} (b)}}
\caption[]
{\label{fig2} \noindent
(a) LEDC unit with excess electron on left-most carbonyl carbon, bent
out of the plane formed by its three oxygen neighbors; (b) LEDC without
an excess electron.
}
\end{figure}

To estimate the reduction thresholds, we take a single formula unit of
Li$_2$CO$_3$ (Fig.~\ref{fig1}a) and LEDC (Fig.~\ref{fig1}b), embed each into
the SMD dielectric continuum,\cite{smd} and perform cluster-based DFT/PBE0
calculations to evaluate the free energy difference between LEDC and LEDC$^-$,
and between Li$_2$CO$_3$ and Li$_2$CO$_3^-$.  These calculations assume a high,
liquid-electrolyte-like dielectric constant ($\epsilon_o$), are pertinent
to SEI products in liquid or at liquid-SEI interfaces, and likely overestimate
$\Phi$ and the stability of SEI products inside the lower dielectric regions
of solid SEI phases, for which crystal structures are frequently lacking.

By manually deforming one of the two CO$_3$ groups in LEDC into a tetrahedral,
$sp^3$-like geometry, we manage to deposit an excess $e^-$ on that deformed
CO$_3$ group (Fig.~\ref{fig2}a).  The optimized configuration is substantially
distorted from the initial optimized, unreduced LEDC geometry
(Fig.~\ref{fig2}b).  It has
O-C-O angles of 109.7$^o$, 114.4$^o$, and 116.6$^o$, and more significantly,
the C-atom is out of the plane formed by the three O atoms by 0.35~\AA\,
instead of being co-planar in the uncharged case.  The predicted $\Phi$ is
$-0.44$~V vs.~Li$^+$/Li(s), which is outside the operating window of lithium
ion batteries.  Attempts to inject an $e^-$ into the CO$_3^{2-}$
unit\cite{li2co3_dftu} invariably leads to one of the Li$^+$ ion being reduced
to a charge-neutral Li radical instead, along with a negative $\Phi$=$-0.40$~V
vs.~Li$^+$/Li(s).\cite{note1,qi2}  We conclude that both LEDC and
Li$_2$CO$_3$ are electrochemically stable under normal battery operating
conditions.  For either to accept $e^-$ from the anode, reactions triggered
by the interface or other chemical means are required.

\subsection{Interfacial Instability: LEDC on Li(100)}

\begin{figure}
\centerline{\hbox{ \epsfxsize=1.5in \epsfbox{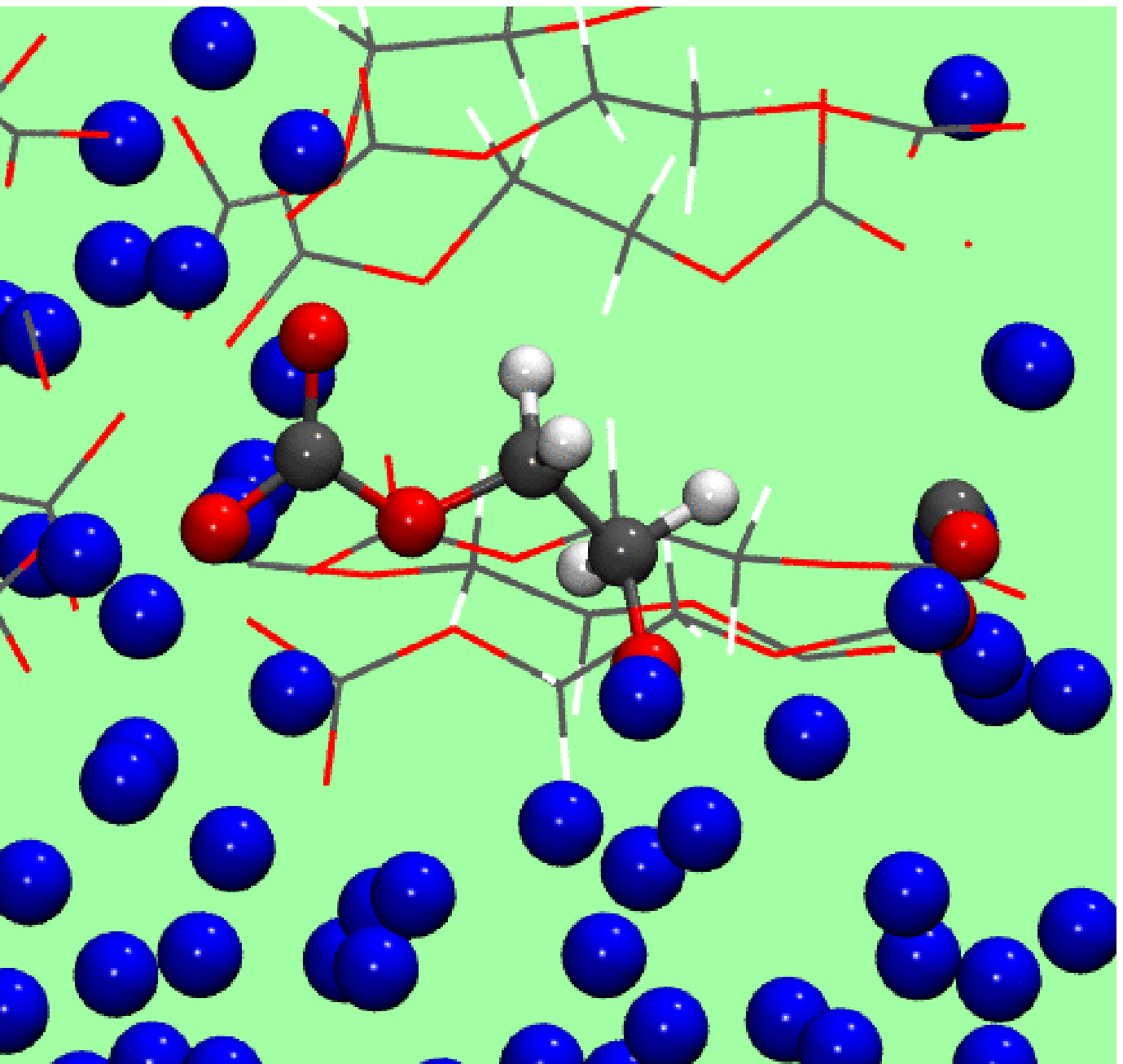}
                   \epsfxsize=1.5in \epsfbox{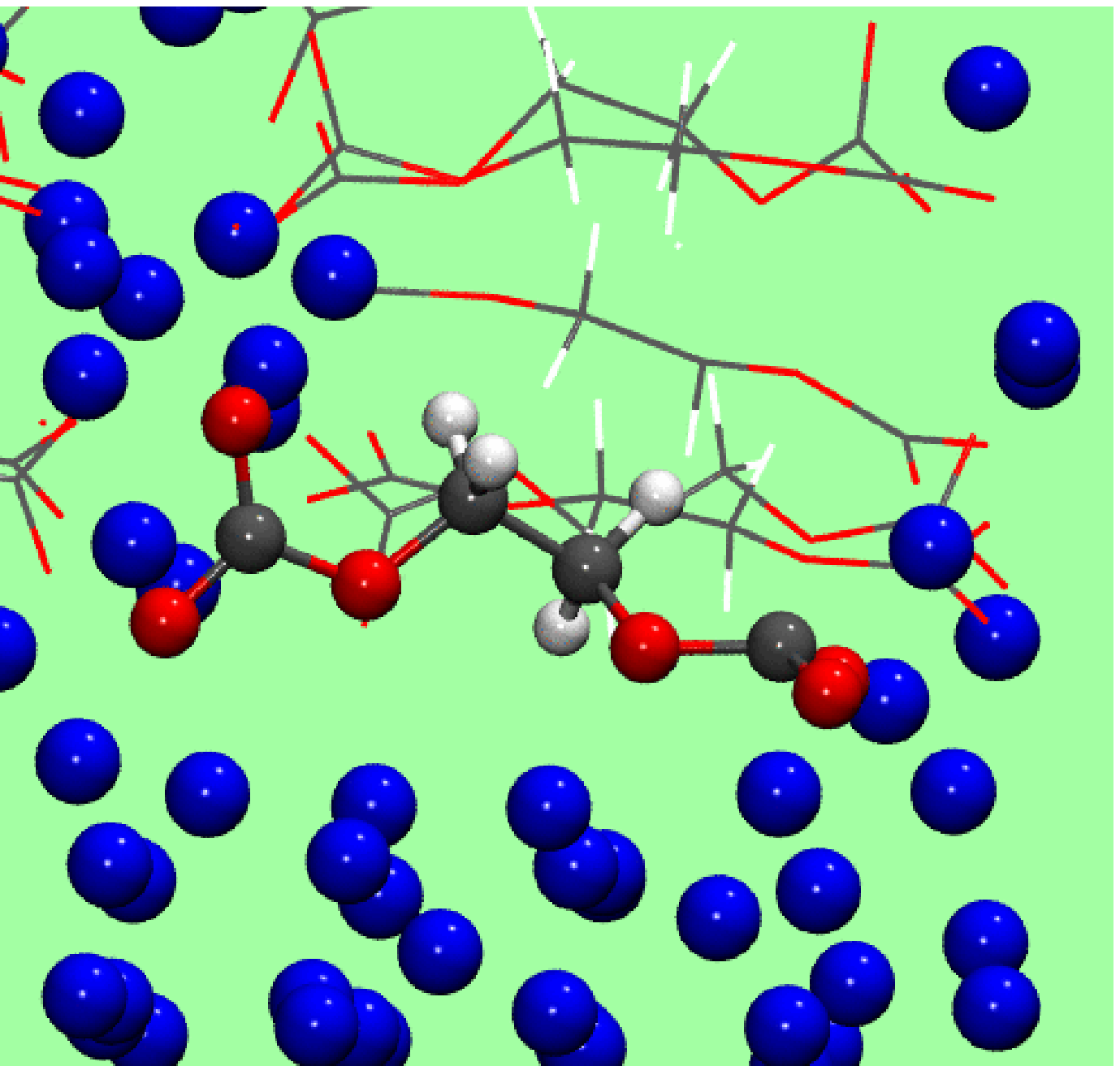} }}
\centerline{\hbox{ (a) \hspace*{1.29in} (b) }}
\caption[]
{\label{fig3} \noindent
(a) LEDC on Li(100) with broken C-O bond and CO$_2$ unit on Li surface; (b)
reaction transition state.  See Fig.~\ref{fig1}c, which depicts the unreacted
interface, for color key.
}
\end{figure}

\begin{figure}
\centerline{\hbox{ \epsfxsize=4.5in \epsfbox{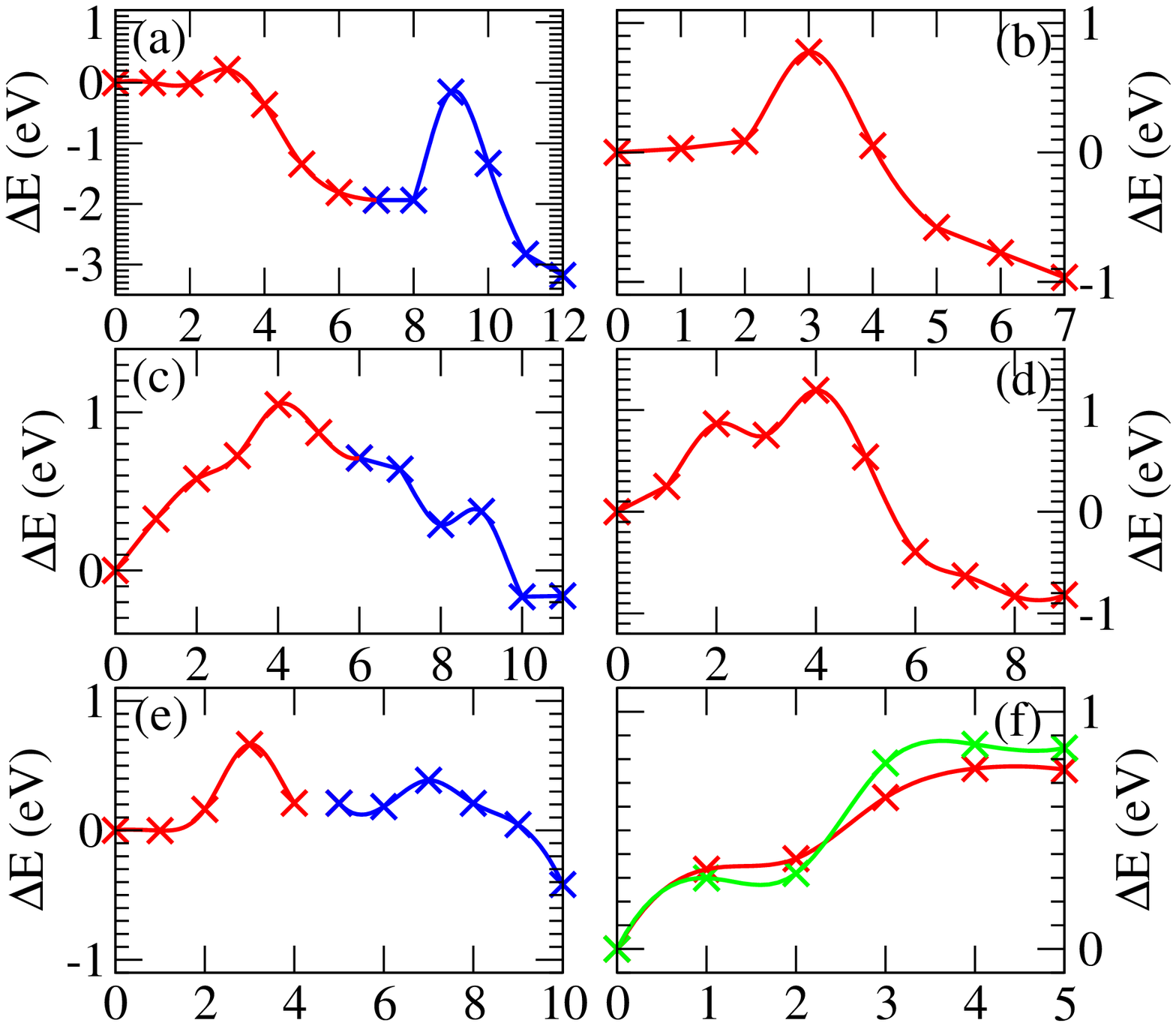}}}
\caption[]
{\label{fig4} \noindent
NEB reaction energy profiles, illustrating reaction barriers and
exothermicities.  (a) LEDC on Li(100), red and blue denote the detachment of
first a CO$_2^{2-}$ and then a O$^{2-}$; (b) CO$_2^{2-}$ reaction on Li(100);
(c) LEDC on a-Si, red and blue denote formation of C-Si bond and then breaking
of C-O bond; (d) Li$_2$CO$_3$ decomposition on Li(100); (e) a lower barrier
Li$_2$CO$_3$ reaction on Li(100) surface with 2~extra Li inserted at the
interface, with Li$_2$CO$_3$ bending and C-O bond-breaking depicted in red
and blue, respectively; (f) out-of-plane bending of a single Li$_2$CO$_3$ unit
on Li(100), where red and green denote PBE and PBE0 predictions.  In panels
(c) and (d), the final barrier heights are obtained by quasi-Newton, non-NEB
optimization of approximate barrier-top configurations generated from
almost-converged NEB runs.
}
\end{figure}

In the remainder of this section, we focus on interfacial studies.
Figure~\ref{fig1}d depicts a 3-bilayer LEDC slab on the Li(100) surface.  We
use the climbing-image NEB technique to examine the reaction energetics of a
EDC with one -CH$_2$O(C=O)O$^-$ group close to and parallel to the Li metal
surface.  The reaction end product, with a broken C-O bond and a released
CO$_2^{2-}$ which binds to the Li surface (Fig.~\ref{fig3}a),\cite{co2,co2_ref}
is exothermic by 1.94~eV (Fig.~\ref{fig4}a).  The DFT/PBE-predicted reaction
barrier of $\Delta E^*$=0.22~eV associated with the transition state
(Fig.~\ref{fig3}b) is low enough to permit sub-hourly reactions.  While the
PBE functional can slightly underestimate reaction barriers, this small
$\Delta E^*$ means that using a more accurate DFT functional is unlikely to
change our qualitative conclusion about the reaction time scale. 

The electronic voltages of this interface with/without LEDC decomposition,
computed using the work function approach,\cite{solid} are $-0.08$~V and
$-0.06~V$ vs.~Li$^+$/Li(s) before and after C-O bond cleavage.  They are only
slightly below the Li-intercalation voltage for Si.  Only small changes in
the potential accompany in this charge-transfer reaction, unlike in
Ref.~\onlinecite{otani12}.  This is partly because the voltage change after
charge transfer is inversely proportional to the simulation cell surface area,
which is much larger in our system (Table~\ref{table1}).  Our attempt to
increase the voltage by removing Li$^+$ at interfaces and creating favorable
dipole moments\cite{solid} leads to diffusion of Li$^+$ from the outer region
to the inner region that negates the intended, initially observed voltage
increase.  Thus we have omitted voltage control for LEDC films (see however
Li$_2$CO$_3$ below).

Soto {\it et al.} have also reported EDC decomposition in unconstrained,
picosecond-long AIMD simulations, including CO$_2^{2-}$ and CO$_3^{2-}$
detachment from LEDC, in the absence or presence of radicals in the
solution.\cite{soto}  That work was performed on 2 layers of LEDC on Li-rich
Li$_x$Si surfaces.  This suggests that interfacial chemical reactions are not
affected by the thickness of the LEDC film or the precise chemical/dielectric
medium in which the EDC unit resides.  A recent UHV measurement has shown
that LEDC can be stabilized on lithium metal surfaces at T=100~K.\cite{robey}
The voltage under UHV conditions should be close to the bare Li(100) value
of 1.56~V.\cite{solid}  In light of the present work, SEI decomposition on
lithium surfaces at T$>$100~$^o$C temperature in UHV conditions, with
an applied electric field to mimic low-voltage conditions, may be of
significant interest.  Finally, reactions involving
oligomerization\cite{tavassol,soto} or proton transfer\cite{shkrob} may occur,
but the CO$_2^{2-}$ release route is already sufficiently fast to ensure that
decomposition takes place.  For the same reason, we have not examined the
effect of LEDC orientation on its decomposition pathways on Li(100).

The CO$_2^{2-}$ fragment on Li(100) surfaces
can further react.  The barrier associated with breaking one of its C-O bonds
($\Delta E^*$=0.78 eV) is somewhat higher than the first EDC decomposition
step.  The reaction is also less exothermic ($\Delta E$=$-0.97$~eV,
Fig.~\ref{fig4}b), but should readily proceed within battery operation time
scales.  It is unnecessary to compute the barrier associated with CO reactions
on Li(100); in a previous AIMD simulation, CO breakdown into C and O dispersed
inside Li metal has been observed within picoseconds.\cite{bal_collab} The role
of CO$_2$ in improving the anode SEI has been discussed in the
literature.\cite{dudney,osaka}

Finally, we examine the interfacial stability of the alkoxide terminus
(-CH$_2$O$^-$) coordinated to Li metal surface, left over from the release
of CO$_2$ from EDC.  While the reaction is exothermic, the barrier
is almost 1.8~eV (Fig.~\ref{fig4}a).  From this calculation, we infer that
\begin{equation}
{\rm OCH}_2{\rm CH}_2{\rm O}^{2-} + 4 {\rm Li} \rightarrow 
	2 {\rm Li}_2{\rm O} + {\rm C}_2{\rm H}_4 ({\rm g}) \label{eq3}
\end{equation}
will not proceed on Li metal surfaces on timescales relevant
to battery operations.  X-ray photoelectron spectroscopy measurements 
have reported the existence of alkoxide groups near lithium metal
surfaces,\cite{aurbach93,aurbach99} and Li-C bonds have also been
reported.\cite{aurbach93}  Our calculations suggest the latter may arise from
breakdown of CO$_2^{2-}$ fragments. Some alkoxide groups may further react
with organic carbonate molecules.\cite{e2,gachot}

In summary, LEDC is predicted to decompose into O$^{2-}$, carbon (either
as lithium carbide or graphite), and OC$_2$H$_4$O$^{2-}$ on Li(100)
surfaces.  

\subsection{Interfacial Stability: LEDC on a-Si}

\begin{figure}
\centerline{\hbox{ \epsfxsize=1.48in \epsfbox{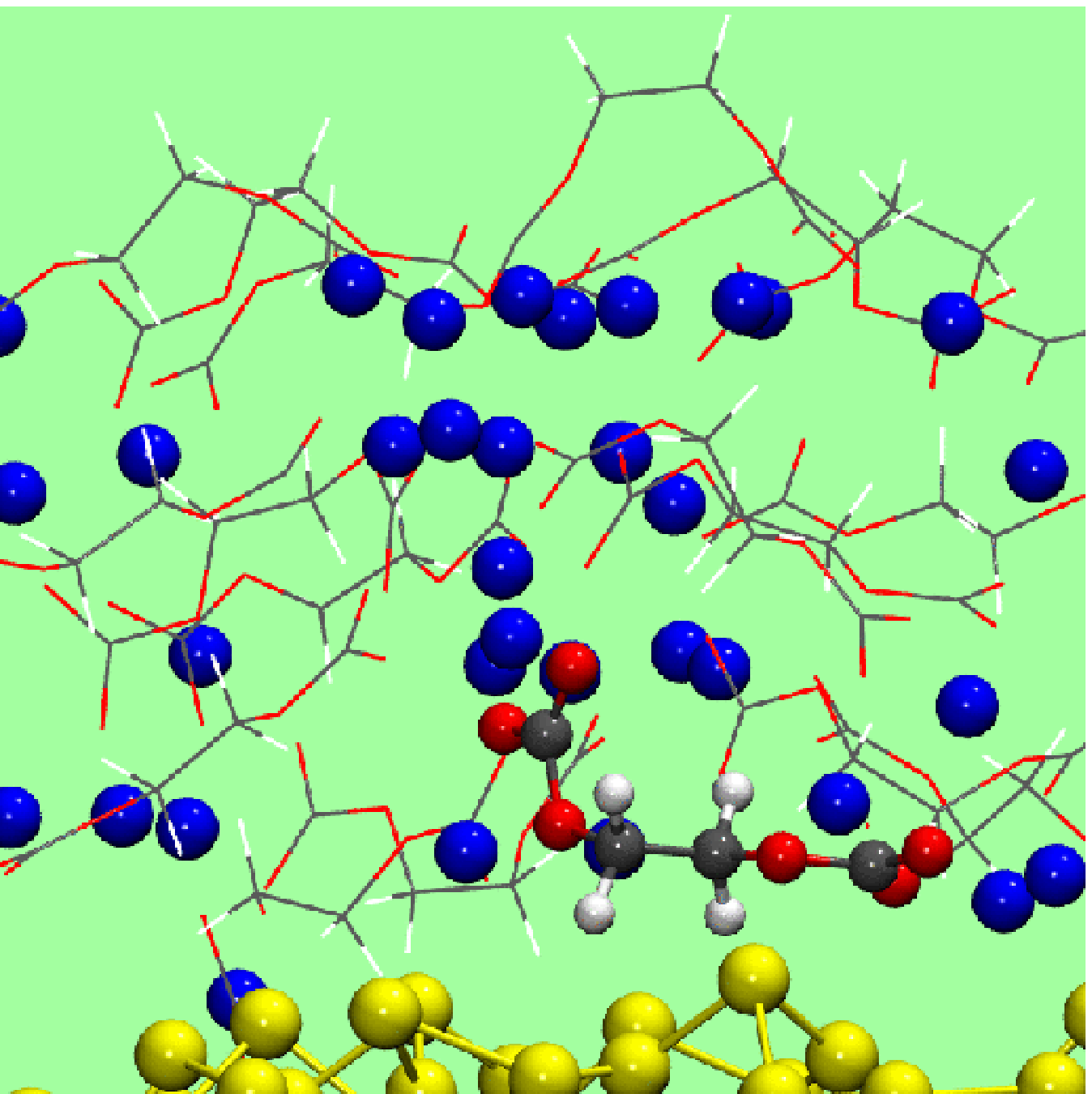}
                   \epsfxsize=1.48in \epsfbox{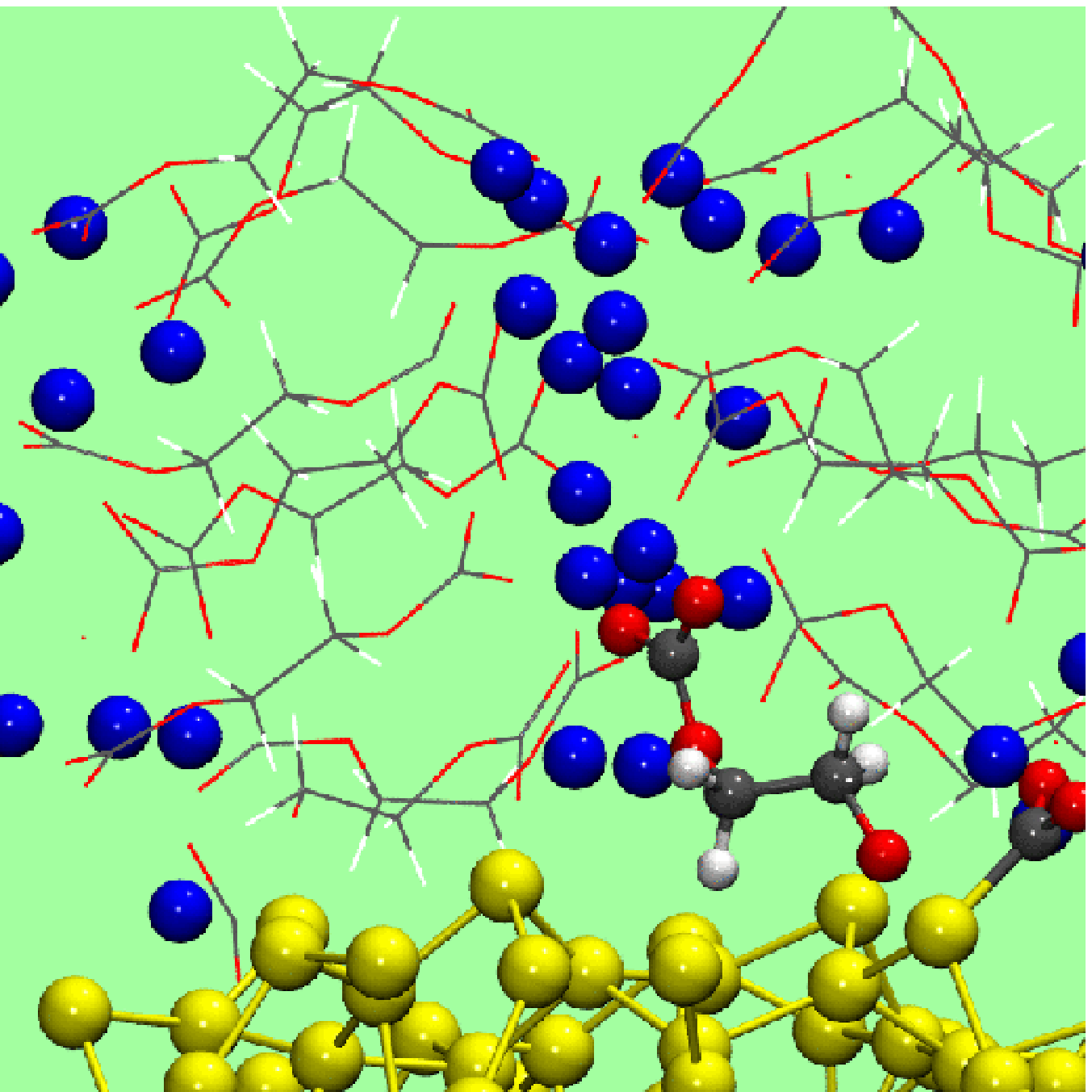} }}
\centerline{\hbox{ (a) \hspace*{1.29in} (b) }}
\caption[]
{\label{fig7} \noindent
(a) LEDC on a-Si surface.  (b) C-O bond cleavage.
}
\end{figure}

Figure~\ref{fig7}a depicts a 3-bilayer LEDC film on a-Si surface.  The
interface model originates from LEDC adsorbed on LiSi. Li atoms are sequentially
removed from LiSi and the lateral dimensions of the simulation cell are
contracted in stages.  Due to the contraction, EDC appear buckled with their
ionic termini bent towards the surface.  Motivated by the predictions on
Li-rich surfaces, we attempt to break a C-O bond on a EDC molecule to release
a CO$_2$ group which becomes bonded to a surface Si atom via its C-site
(Fig.~\ref{fig7}b).  (This LEDC molecule chosen to react has one carbonyl~C
atom closest to an Si atom, and should react most readily among those adsorbed
on the surface.)  The reaction is slightly exothermic at zero temperature
($\Delta E$=-0.17~eV), and the overall barrier is $\Delta E^*$=0.95~eV
(Fig.~\ref{fig4}d), slightly higher than the $<0.92$~eV kinetic criterion
discussed in the Method section.  This shows that LEDC is kinetically
barely stable on discharged silicon anode surfaces.  It cannot be ruled out
that other starting configurations on the amorphous Si surface may yield
slightly lower reaction barriers, but LEDC reactivity is clearly much
lower on a-Si than Li(100).

\subsection{Interfacial Instability: Li$_2$CO$_3$ on Li(100)}

\begin{figure}
\centerline{\hbox{ \epsfxsize=1.5in \epsfbox{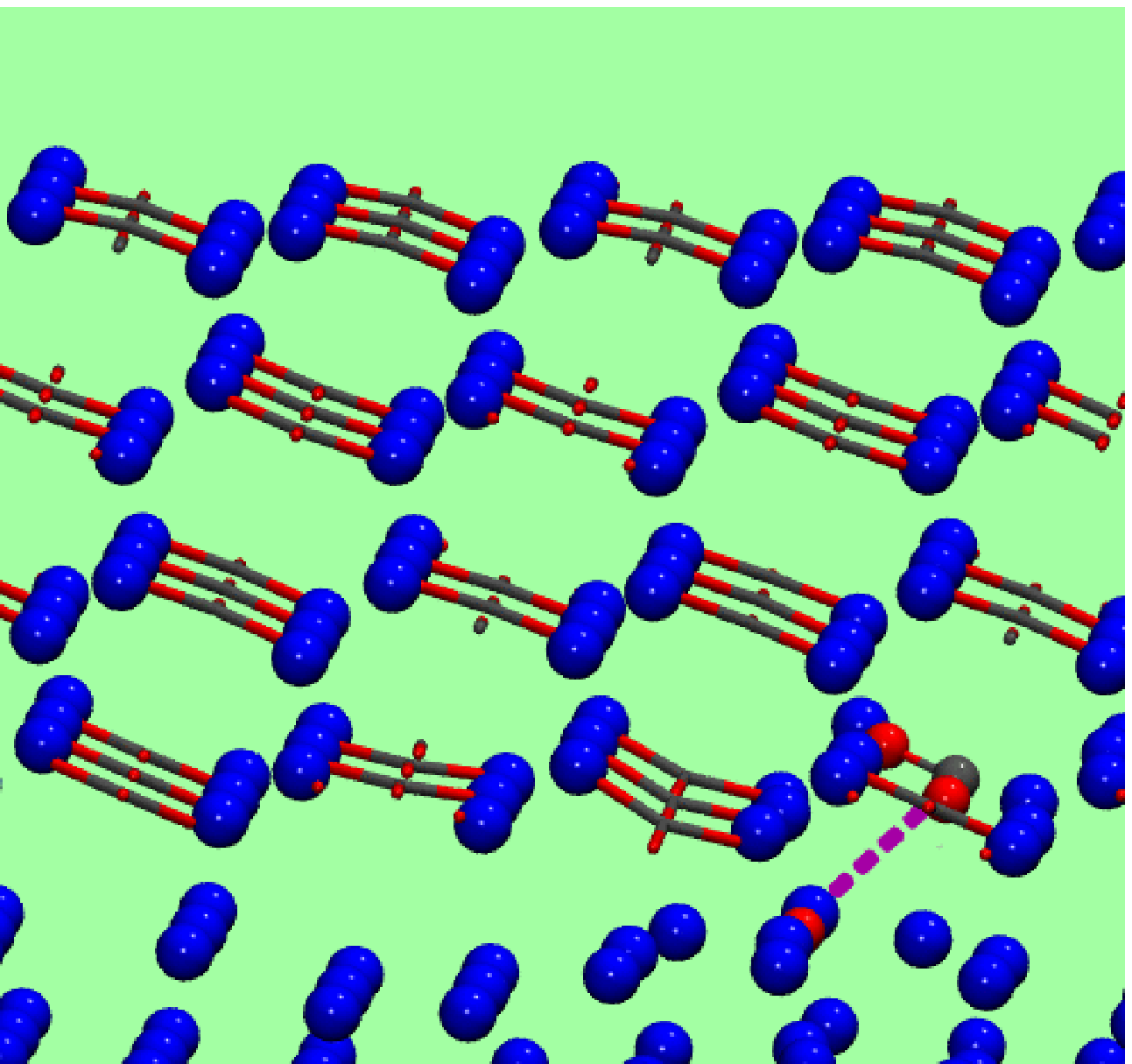}
                   \epsfxsize=1.5in \epsfbox{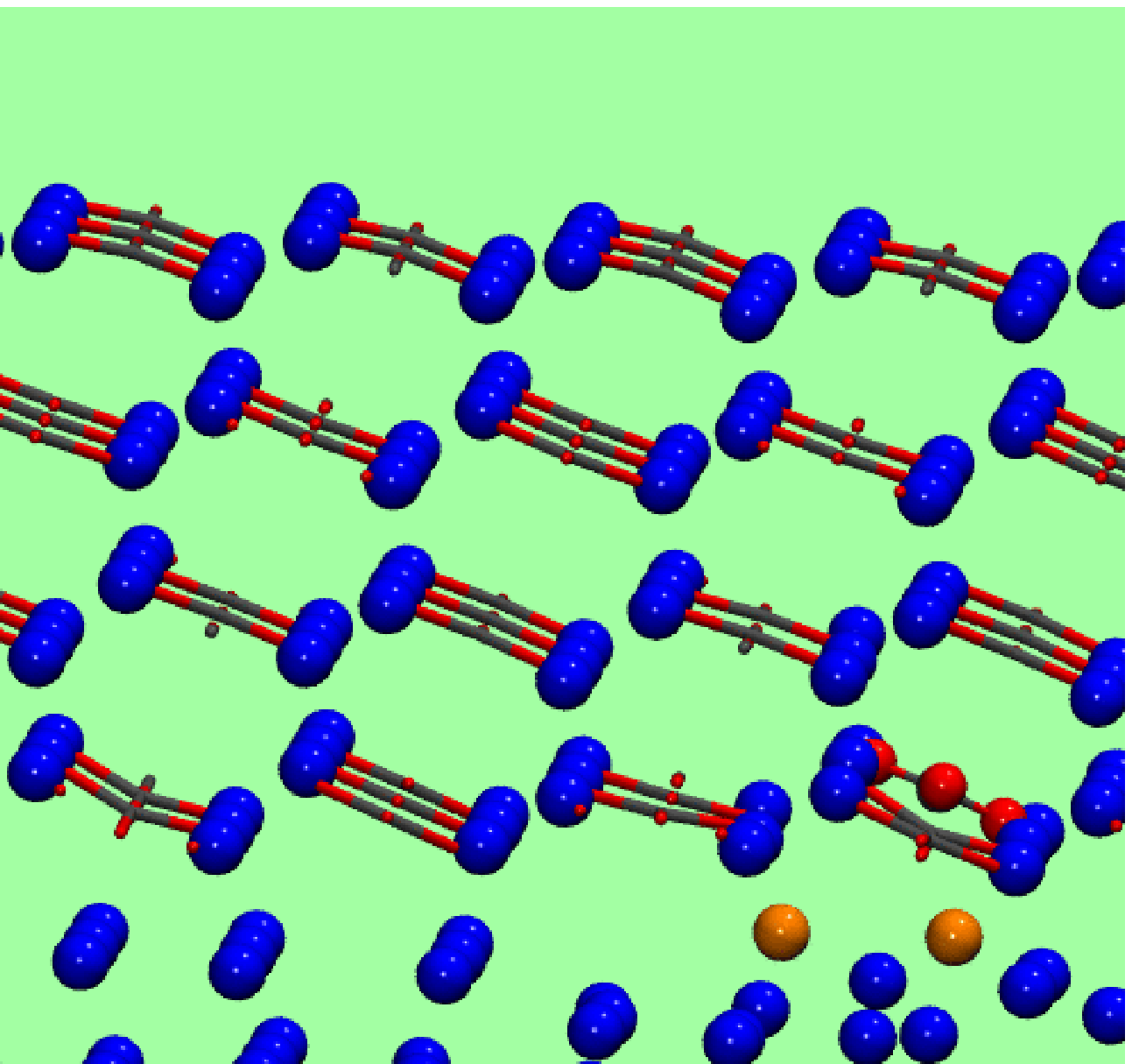} }}
\centerline{\hbox{ (a) \hspace*{1.29in} (b) }}
\centerline{\hbox{ \epsfxsize=1.5in \epsfbox{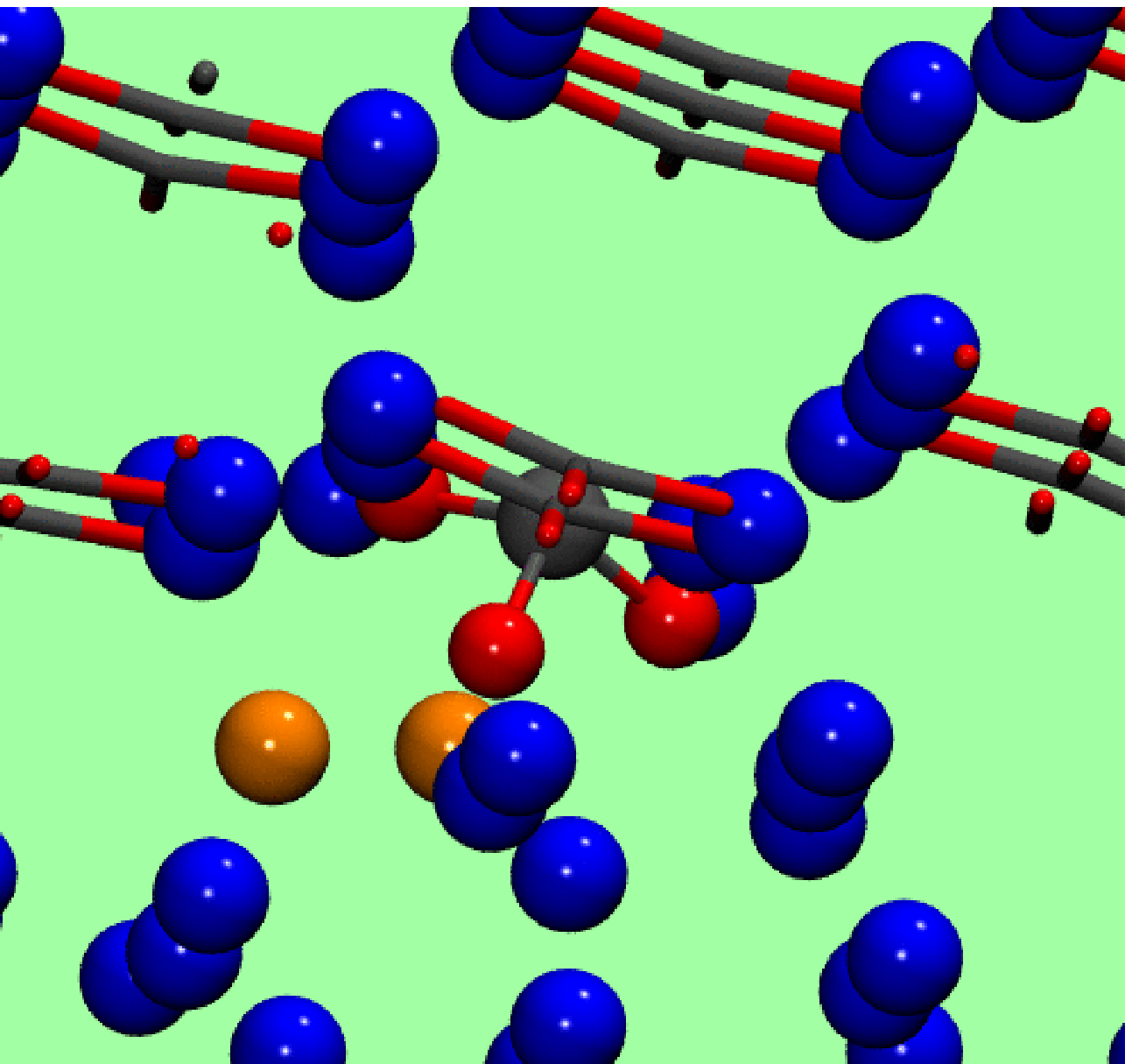}
                   \epsfxsize=1.5in \epsfbox{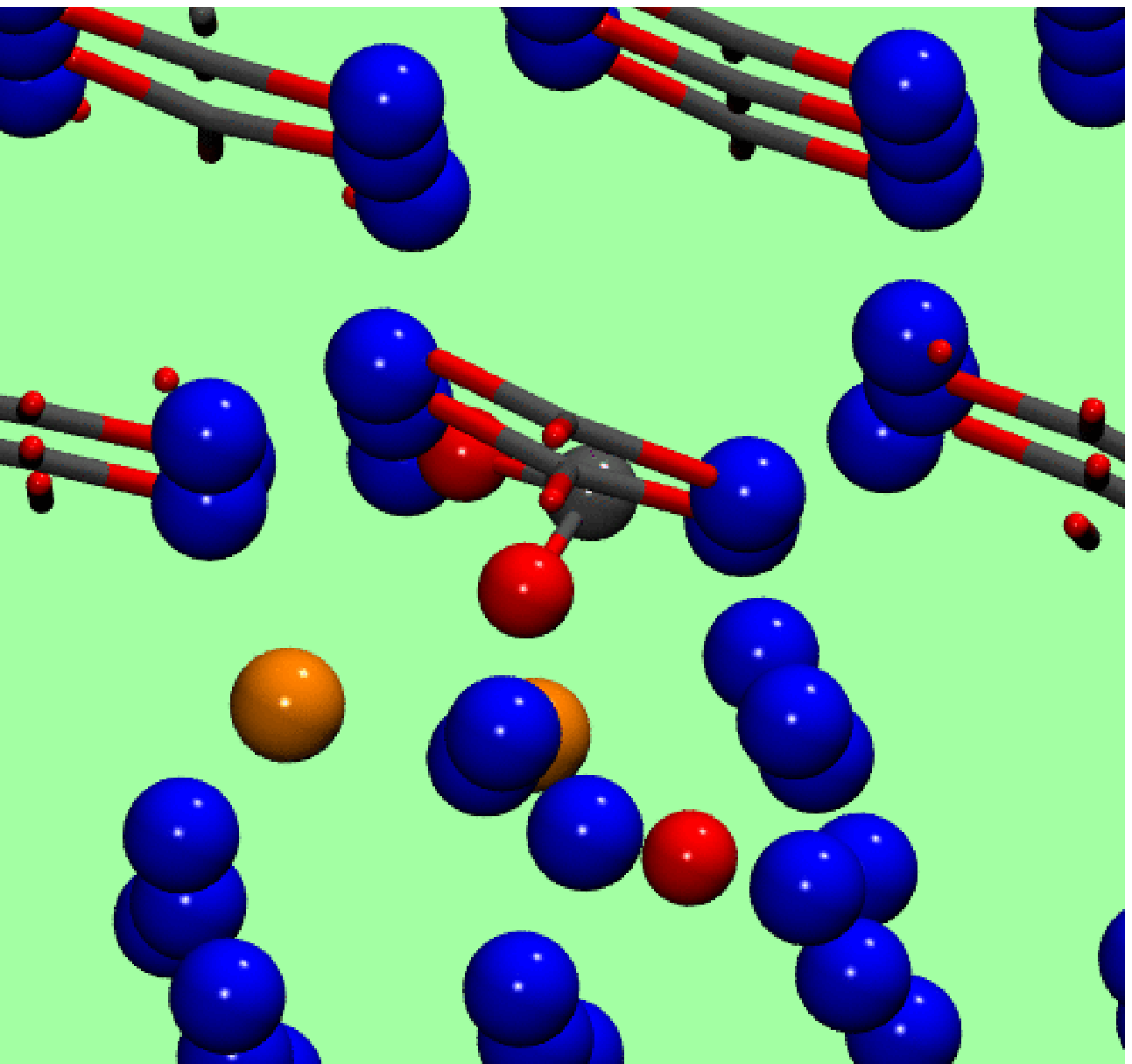} }}
\centerline{\hbox{ (c) \hspace*{1.29in} (d) }}
\centerline{\hbox{ \epsfxsize=1.5in \epsfbox{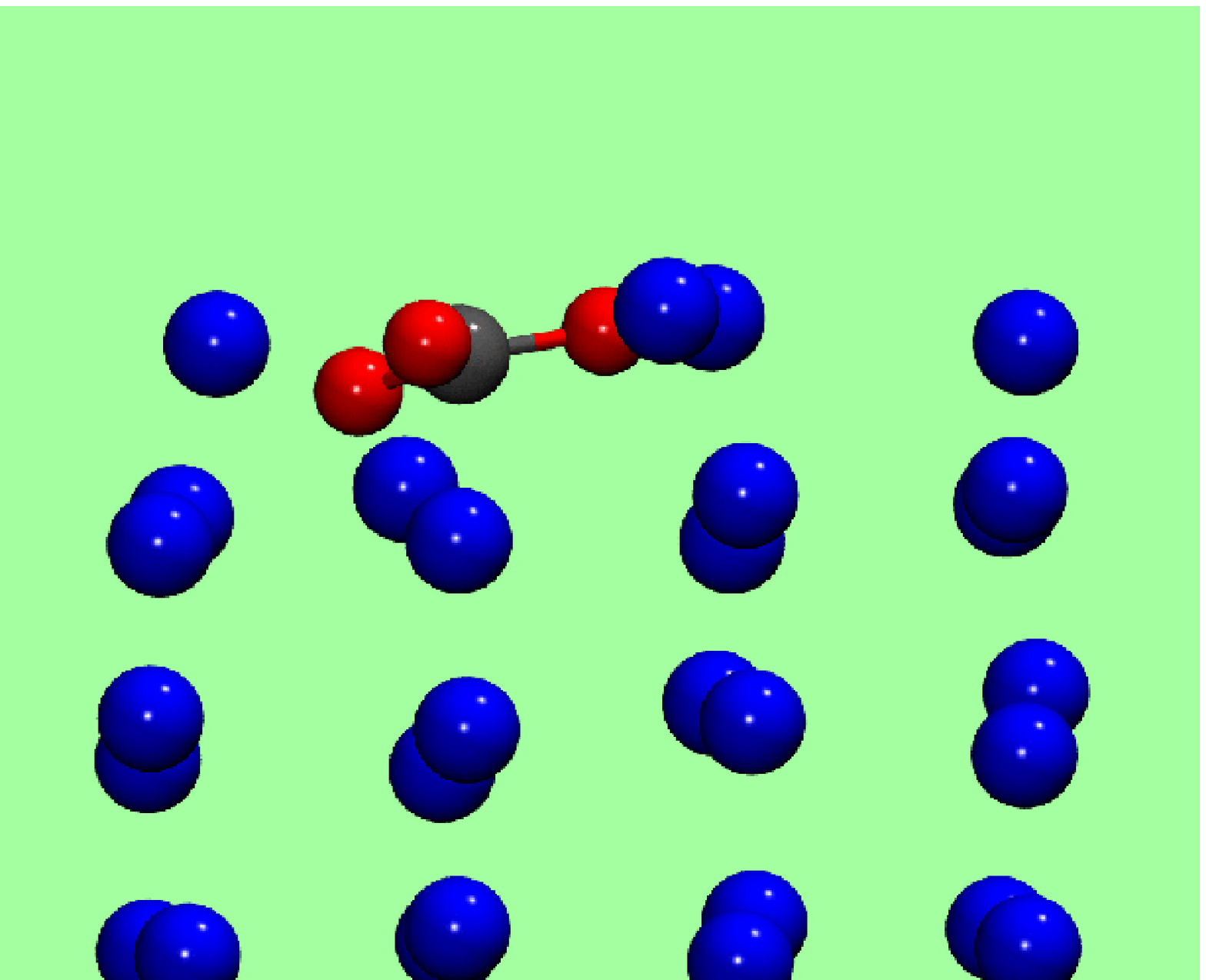}
                   \epsfxsize=1.5in \epsfbox{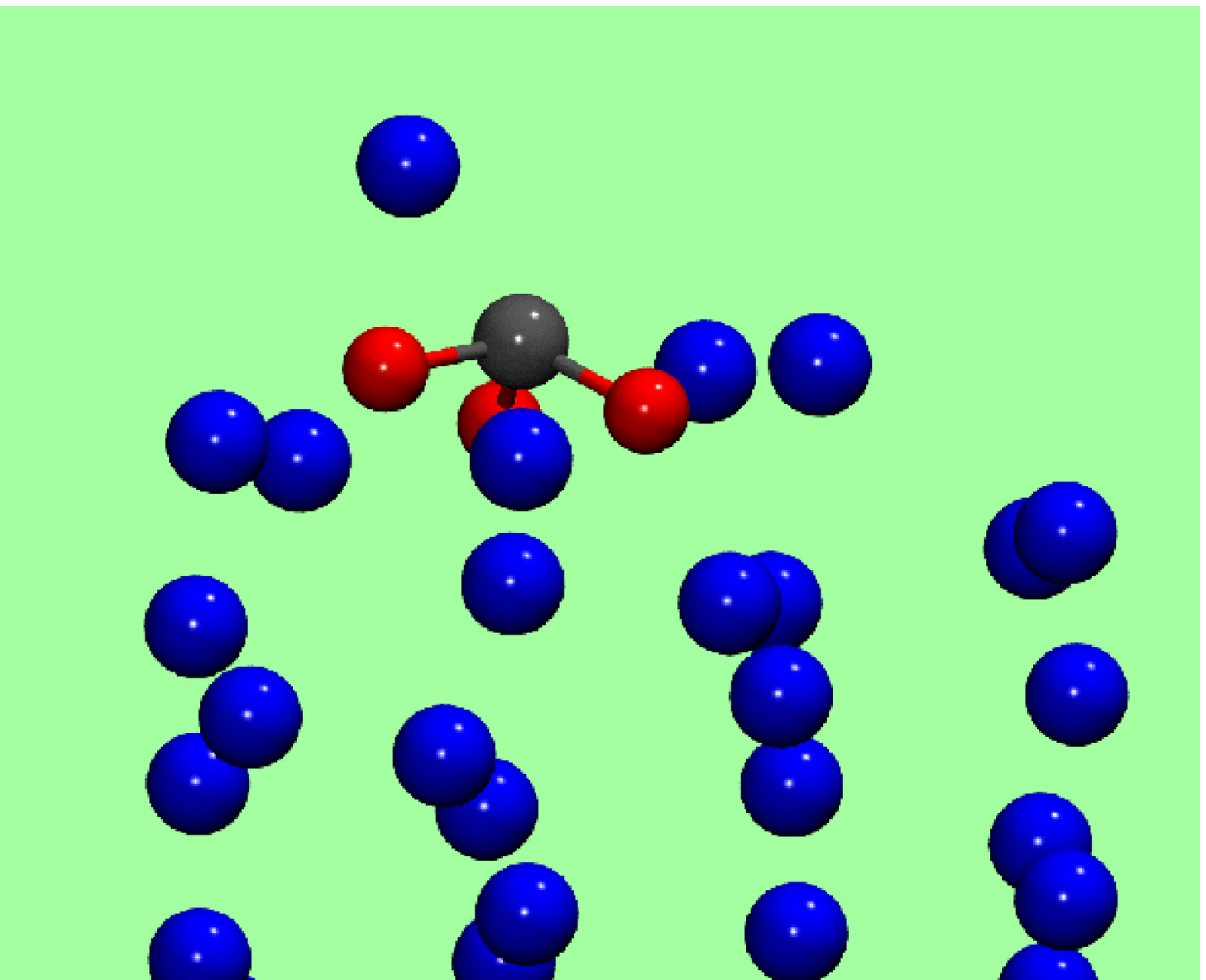} }}
\centerline{\hbox{ (e) \hspace*{1.29in} (f) }}
\centerline{\hbox{ \epsfxsize=1.5in \epsfbox{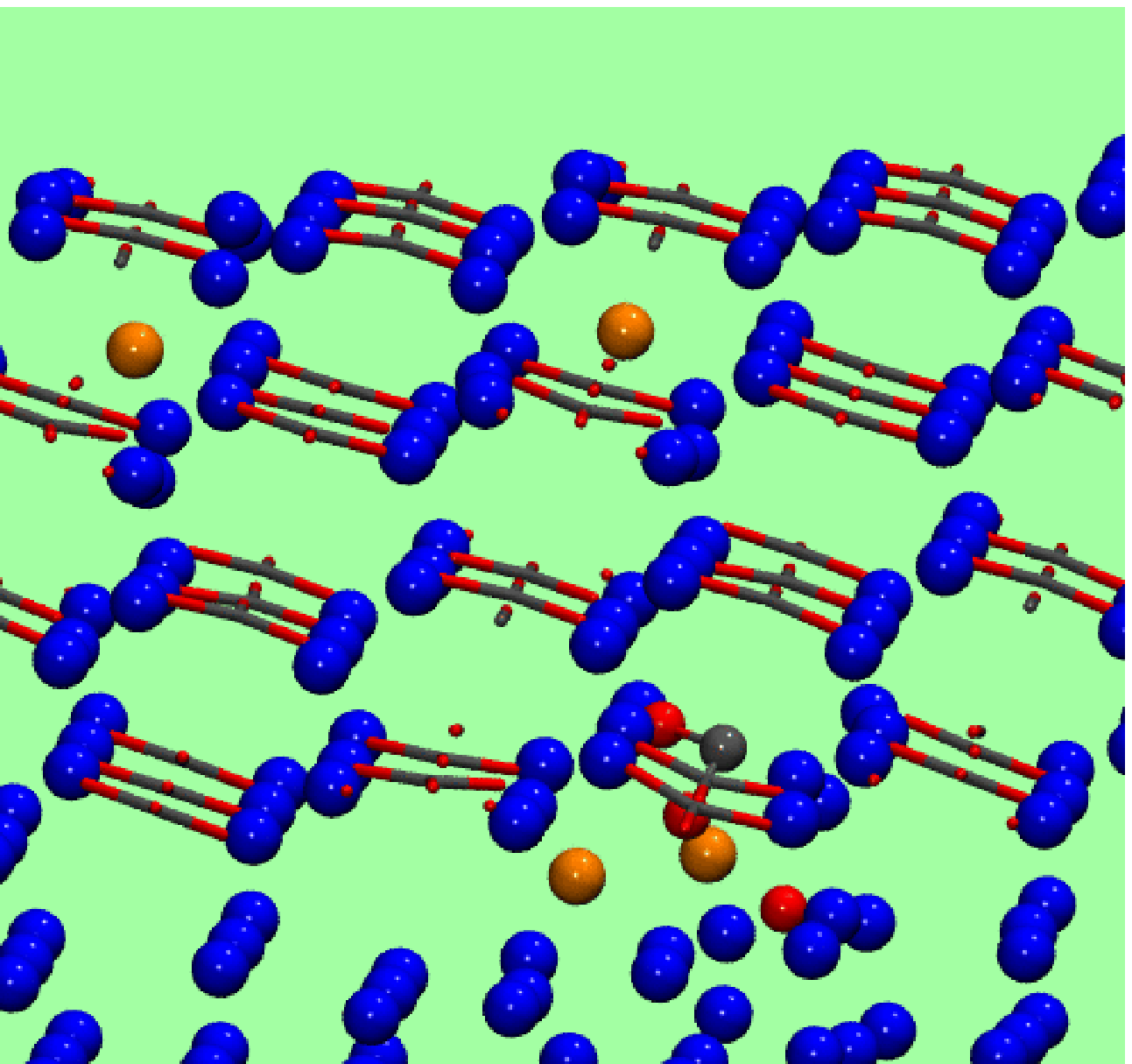}
                   \epsfxsize=1.5in \epsfbox{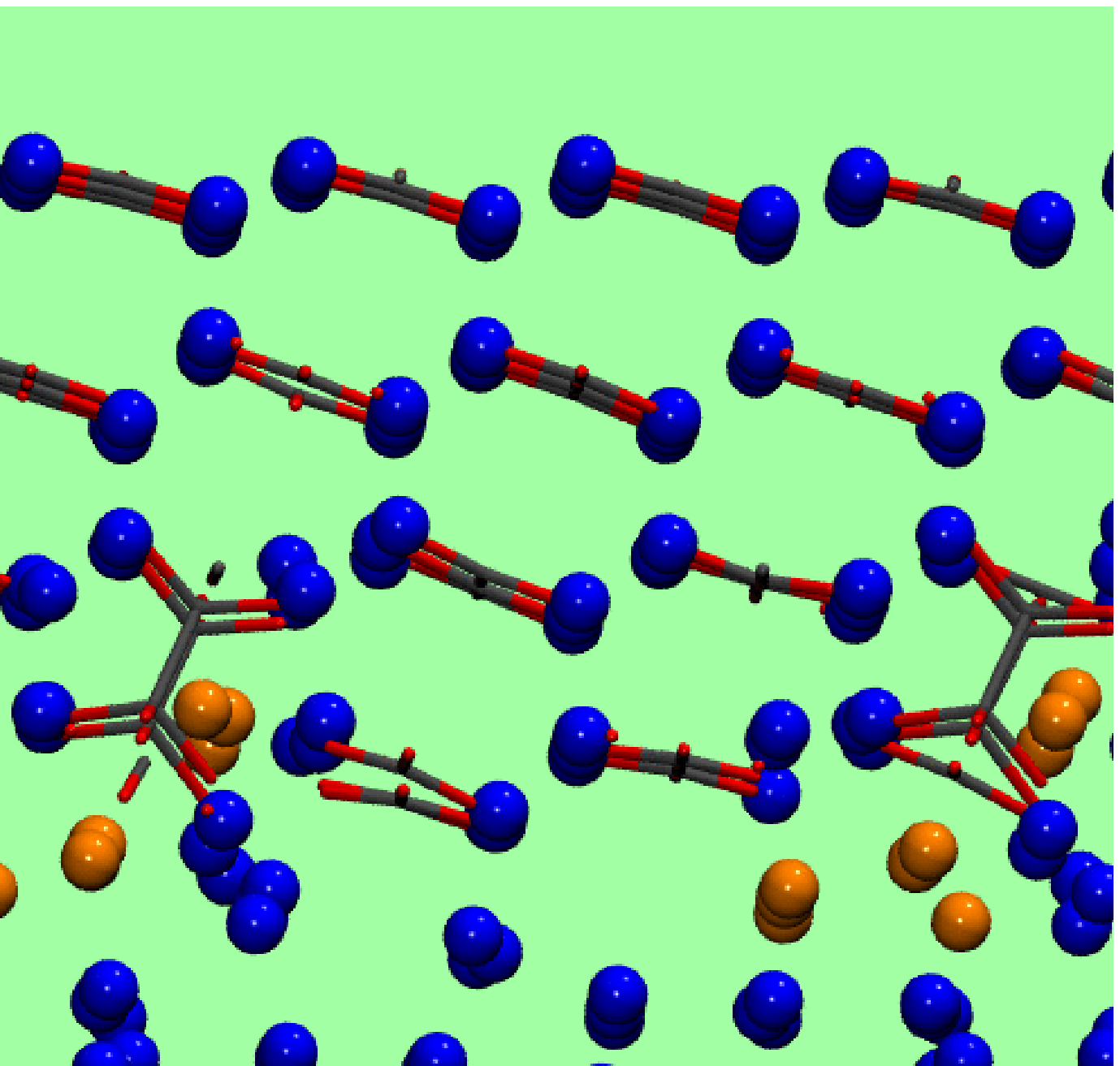} }}
\centerline{\hbox{ (g) \hspace*{1.29in} (h) }}
\caption[]
{\label{fig8} \noindent
(a) Li$_2$CO$_3$ (001) on Li(100), with broken C-O bond (red line) at interface;
(b) intact Li$_2$CO$_3$ but with 2 extra Li at interface; (c) reaction
intermediate with C-atom out of plane; (d) final configuration with broken
C-O bond.  (e) A single CO$_3^{2-}$ on Li(100); (f) bent geometry intermediate.
(g) Same as panel (d), but with two Li$^+$ interstitials between the top-most
LiCO$_3$ layers.  (h) Spontaneous reactions forming 4 C-C bonds when 12
extra Li are added at interface, and 6 Li are inserted as interstitials between
the Li$_2$CO$_3$ layers closest to the Li surface.  Extra Li are shown
in orange.
}
\end{figure}

Li$_2$CO$_3$ is thermodynamically unstable at voltages near Li-plating when
excess Li is available (Eq.~\ref{eq1}).  The only question is whether a
kinetically viable pathway allows this and related reactions.
Figures~\ref{fig8} depicts Li$_2$CO$_3$ decomposition on Li(100).  Initially
we place a 4-layer Li$_2$CO$_3$ (001) slab on Li(100) and optimize the geometry
(Fig.~\ref{fig1}e).  Substantial deformation of the soft lithium metal surface
accommodates the presence of the carbonate film.\cite{note2}  Breaking a
C-O bond at the interface, leaving a CO$_2^{2-}$ anion in the carbonate layer
while depositing an oxygen anion on the Li surface (Fig.~\ref{fig8}a), is
exothermic by $-0.82$~eV.  The overall transition state barrier energy
(Fig.~\ref{fig4}d) is 1.19~eV, meaning that this reaction is too slow to occur
during battery operations.  

However, CO$_3^{2-}$ decomposition can be facilitated by excess Li.  When 
two Li atoms are added in what appears to be an empty crevice between Li(100)
and Li$_2$CO$_3$(001) (Fig.~\ref{fig8}b), the total energy of the simulation
cell is only 0.14~eV higher after subtracting the chemical potentials
($\mu_{\rm Li}$) of the two added Li, which are assumed to be the
$\mu_{\rm Li}$ of Li metal.  At higher equilibrium potentials, the cost
will be larger, which slightly increases the overall barrier.  Configurations
with excess Li at the interface may be inevitably present during charging,
because Li has to pass through the interface.

Figures~\ref{fig8}b-\ref{fig8}d revisit CO$_3^{2-}$ decomposition when these
two Li are present.  In contrast to the case without additional Li atoms,
a metastable reaction intermediate can be stabilized.  It consists of
the decomposing CO$_3^{2-}$ adopting a bent geometry, with excess $e^-$ on the
C atom (Fig.~\ref{fig8}c).  The rate-determining barrier to reach this
intermediate is $\Delta E^*$=0.66~eV (Fig.~\ref{fig4}e), much reduced from
the case without the two additional interfacial Li (Fig.~\ref{fig4}d).  
The further reaction to break a C-O bond in the bent CO$_3^{2-}$ and
release O$^{2-}$ is exothermic, with the energy released exceeding the
cost of initially adding the two Li.  This second reaction step exhibits
a minimal barrier (Fig.~\ref{fig4}e).\cite{shenoy}
The two added Li apparently make the reaction zone more $e^-$ rich and 
stabilize the CO$_2^{2-}$ product.  In fact, by placing 18~extra Li atoms
into interfacial crevices (not shown), we have observed several spontaneous,
barrierless CO$_3^{2-}$ $\rightarrow$ CO$_2^{2-}$ + O$^{2-}$ reactions there.
However, the resulting total energy is less favorable than without those
Li extra atoms when $\mu_{\rm Li}$ is accounted for, likely because
some of the added Li are undercoordinated.

The time scale associated with planar-CO$_3^{2-}$$\rightarrow$bent-CO$_3^{4-}$,
deformation (0.66~eV barrier) is estimated to be 0.12~s, within battery
operation timescale but far beyond AIMD trajectory lengths.  This emphasizes
the need to calculate barriers, using either static NEB calculations or
liquid state potential-of-mean-force techniques as appropriate.  

We also consider a single CO$_3^{2-}$ adsorbed on Li(100) (Fig.~\ref{fig8}e).
This system involves a much smaller simulation cell and permits the use
of more accurate DFT functionals to re-examine the rate-determining barrier
required to reach the deformed CO$_3$ intermediate.  Four Li atoms
decorate the area around the anion so that a bent, metastable CO$_3$
(Fig~\ref{fig8}f) can be stabilized at +0.76~eV vs.~the original flat
CO$_3^{2-}$ geometry.  The barrier leading to this intermediate is 0.76~eV.
Using the generally more accurate DFT/HSE06 method instead of PBE used
throughout this subsection, the barrier is increase only slightly to 0.86~eV.
This demonstrates that there is no significant $\Delta E^*$ dependence on DFT
functionals.  One reason is that this reaction intermediate involves
molecular deformation rather than bond-breaking, and the DFT/PBE
delocalization error should be less significant.\cite{wtyang}  The deformation
energy and electron reduction contributions to the energetics cannot be
decoupled because electrons spontaneously flow to the deformed carbonate.

The electronic voltages of Figs.~\ref{fig8}b-d are all 0.88~V vs.~Li$^+$/Li(s).
It can be argued that the inclusion of liquid electrolytes will reduce this
value to approximately 0~V vs.~Li$^+$/Li(s) (in chemical equilibrium with the
Li metal slab) due to preferential interfacial dipole moment alignment of
organic solvent molecules.\cite{solid,note3}  Since liquid electrolyte is
not considered in this section, we explore the effect of voltage variations by
adding two Li interstitials between the outermost layers of Li$_2$CO$_3$
(Fig.~\ref{fig8}g).  Bader analysis\cite{bader} shows that these Li
spontaneously exhibit Li$^+$ charge states, and their positive charges should
be compensated by a negative surface charge density on Li(100) in the
charge-neutral simulation cell. A large dipole moment is created, lowering the
voltage from 0.88~V to $-0.02$~V.  Even though the overall potential is now
significantly lower than that of Figs.~\ref{fig8}b-d, and this should favors
reductive decomposition, the energies associated with CO$_3^{2-}$ decomposition
are almost unchanged.  The bent CO$_3$ intermediate and the broken C-O
configurations are rendered more favorable only by 0.04 and 0.04~eV,
respectively, compared to when the voltage is 0.88~V (Fig.~\ref{fig8}b-d).
The reason is that the electric field generated by the two Li$^+$ interstitials
is weak ($\sim 0.9$~V/$10$~\AA).  This calculation suggests that, if the outer
surface is occupied by a high-dielectric liquid with counter-ions, so that it
is the electrolyte rather than interstitual Li$^+$ that lowers the electronic
voltage from 0.88~V to about 0~V, the effect on Li$_2$CO$_3$ degradation at the
solid-solid interface will still be small -- just because of the thickness of
the carbonate layer dictates a weak electric field.

The above system with a weak interfacial electric field assumes that the
electric double layer (EDL) is not localized at the solid-solid interface (c.f.
Fig.~1d or Fig.~1f in Ref.~\onlinecite{solid}).  Another possibility is that
Li$^+$ interstitials or vacancies right at the interface lead to a narrow
EDL and a much larger local electric field just outside the active electrode
material (Fig.~1e in Ref.~\onlinecite{solid}).  Without explicit, costly AIMD
free energy simulations of liquid electrolyte outside the SEI film, we cannot
determine which scenario is more likely.  Fig.~\ref{fig8}h explores the
latter possibility by adding 12 Li atoms at the interface and 6 interstitial
Li between the first two carbonate layers closest to the lithium surface.
(Without the added 12 Li at the interface, the interstitial Li$^+$ migrate on
to the Li metal surface.) The charge distribution generates an initial voltage
of about 0.5~V vs Li$^+$/Li(s).  However, upon optimization, C-C linkages
and four (CO$_3$)$_2$ units are formed.  This configuration is more favorable
than Fig.~\ref{fig1}e of this work by 0.08~eV per Li added after accounting for
Li cohesive energy.  It is unclear that a configuration with so many
interstitials is kinetically assessible, and we consider this result
speculative.  Fig.~\ref{fig8}h does emphasize the multitude of reactions
CO$_3^{2-}$ can undergo in this extremely electrochemically reductive
environment, partly due to the large inherent instability associated with
Eq.~\ref{eq1}.  

In the SI, we show that Si atoms substituting for Li on Li(100) surface
reduces reactivity towards Li$_2$CO$_3$.  Li$_2$CO$_3$ decomposition
on an amorphous-Si surface is also shown to be energetically unfavorable.  To
what extent Li$_x$Si, with finite Si content subsurface, will slow down
this reaction will be the subject of future work.

\subsection{Interfacial Instability: AIMD Simulations Including Electrolyte}
\label{aimd}

In this subsection, we consider the initial stages of SEI formation in a less
well-controlled, liquid-solid interface environment,
with LiF decorating the surface.  LEDC instability has been demonstrated
by Soto {\it et al.} on related surfaces.\cite{soto}  Here AIMD simulations
are conducted on a liquid electrolyte and a submonolayer of Li$_2$VDC units
on a Li$_{13}$Si$_4$ surface coated with a 5-\AA\,~diameter LiF cluster
(Fig.~\ref{fig1}g-h).  Li$_{13}$Si$_4$ is chosen because of computational
reasons; its unit cell exhibits better lattice matching with crystalline surface
films.\cite{fec}  The equilibrium voltage of this composition and the maximally
lithiated Li$_{22}$Si$_5$ has been prdicted to differ by less than
0.1~V.\cite{chev}  As discussed in the Introduction, Li$_2$VDC may arise from
vinylene carbonate decomposition, and is structurally similar to LEDC.

\begin{figure}
\centerline{\hbox{ \epsfxsize=4.5in \epsfbox{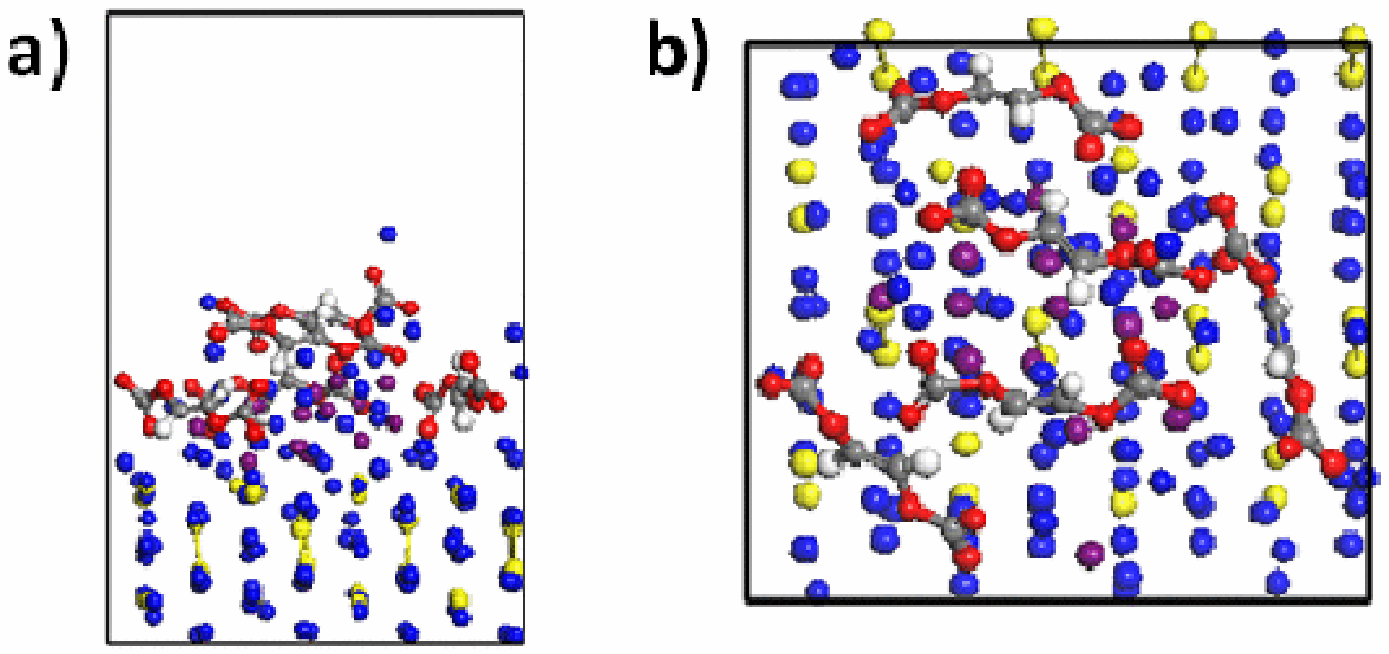} }}
\caption[]
{\label{fig9} \noindent
(a)\&(b) Side and top views of the five Li$_2$VDC oligomers adsorbed at the
surface of the Li-covered lithiated Si anode.  C, O, H, Li, and F atoms are
represented by gray, red, white, blue, and purple spheres, respectively.  The
LiPF$_6$ unit and FEC molecules are removed for clarity.
}
\end{figure}

The outer layer of the thin SEI film is modeled by placing five Li$_2$VDC
oligomers on one side of the slab (Fig.~\ref{fig9}).  Li$_2$VDC are placed
in such a way that the whole surface is covered by the organic block. This
corresponds to a coverage of approximately 1.77 oligomer/nm$^2$ based on the
surface size.  Therefore, our model mirrors an anode with a SEI layer formed
by two components (LiF and Li$_2$VDC) stacked perpendicular to the
Li$_{13}$Si$_4$ (010) surface slab. The liquid electrolyte region of the
battery is placed on top of this surface. The electrolyte consists of pure
FEC solvent molecules and 1.0~M LiPF$_6$ salt.

\begin{figure}
\centerline{\hbox{ \epsfxsize=3.0in \epsfbox{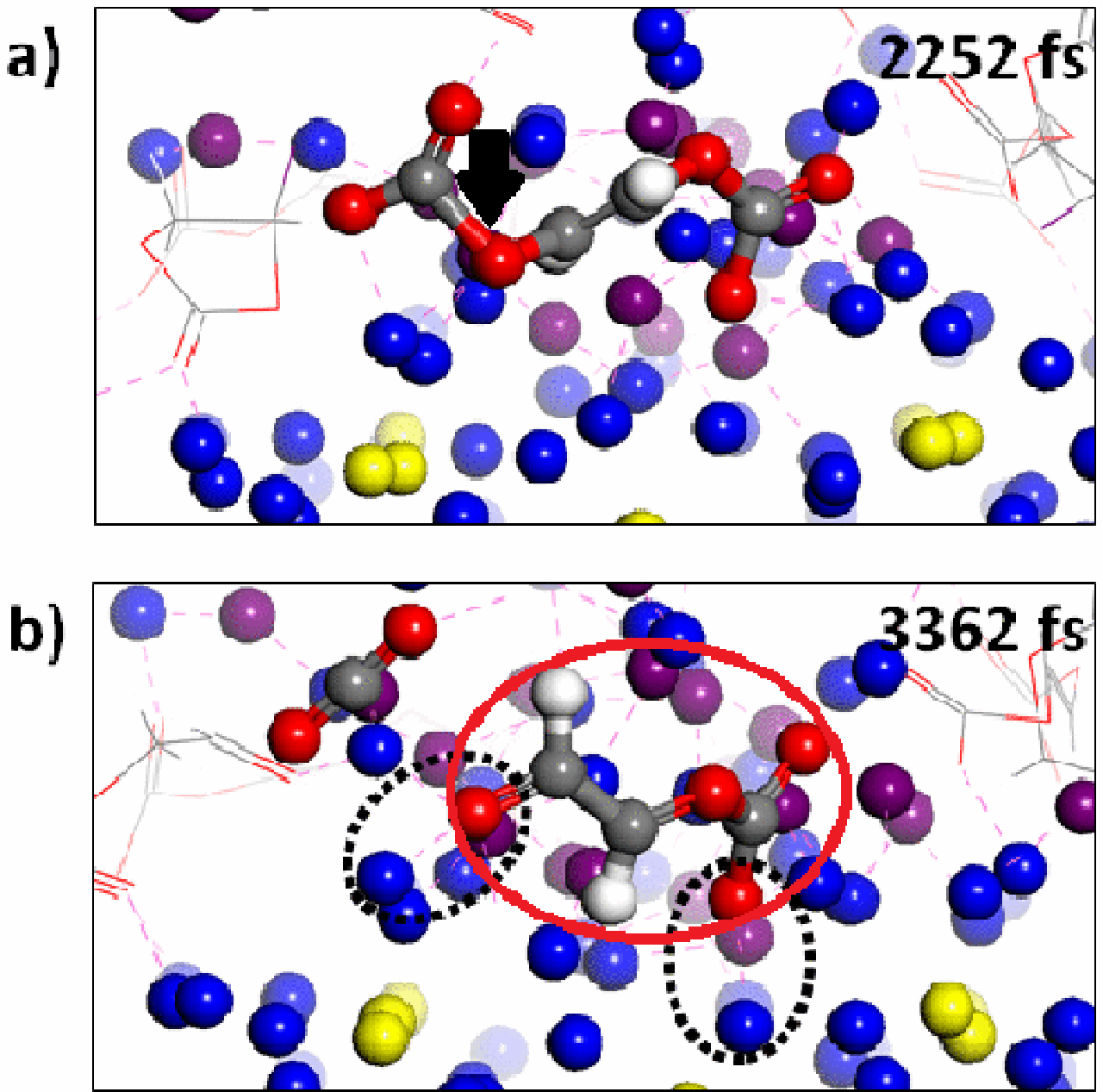} }}
\caption[]
{\label{fig10} \noindent
Zoomed-in snapshots showing the time evolution of the LiF cluster covered
Li$_{13}$Si$_4$ (010) surface in contact with the 1.0 M LiPF$_6$/FEC
electrolyte solution and 5 Li$_2$VDC oligomers adsorbed on the surface
(3 on the exposed area and 2 on top of the LiF cluster). (a) 2.252~ps;
(b) 3.362~ps. Only the dissociating Li$_2$VDC oligomer is shown. FEC
molecules are shown in a line display style. For clarity, a depth cue was
applied to sharply focus on the dissociated oligomer and nearest atoms.
Distant atoms and molecules appear blurrier. 
}
\end{figure}

One Li$_2$VDC molecule in contact with the open lithiated Si surface
is found to decompose.  Starting from the 2.5~ps step, a 
C$_{\rm carbonate}$-O$_{\rm vinylene}$ bond breaks.  The resulting
OC$_2$H$_2$CO$_3^{2-}$ fragment remains adsorbed on the surface,
with the O$_{\rm vinylene}$ atom bonded with two Li$_{\rm surface}$
atoms at an average distance of 1.80~\AA, while the O$_{\rm carbonate}$ atom
is coordinated to a Li$_{\rm surface}$ atom at a distance of 1.83~\AA\,
(Fig.~\ref{fig10}a\& b within the black dashed circles).  A Bader charge
analysis\cite{bader} reveals an almost neutral CO$_2$ molecule, which
diffuses away from the surface, and an OCHCH$_2$ fragment bearing a negative
charge ($-1.79$~$|e|$).  The latter fragment is attached to the CO$_3$
group inside the red cicle in Fig.~\ref{fig10}.  These AIMD simulations
suggest that even Li$_2$VDC oligomers, formed from electrolyte additive
molecules, can be rapidly decomposed when in contact with active Li$_x$Si
anode surfaces decorated with LiF crystallites, in qualitative agreement
with Ref.~\onlinecite{soto}.  Higher Li-content Li$_x$Si are at lower voltage,
and are expected to be even more reactive towards the SEI than Li$_{13}$Si$_4$.

\section{Discussions}
\label{discussions}

The implications of this work are manifold.  First, we stress that
the interfacial reactivities predicted concern reactive anode material
surfaces.  We have not attempted SEI stability analysis on graphite 
interfaces herein.  Even though LEDC and Li$_2$CO$_3$ are always
thermodynamically unstable at low voltages in the presence of excess Li,
interfacial reactions (or other external means) must occur to trigger such
instability.  However, if lithium-plating occurs on the outer surfaces
of SEI-covered graphite,\cite{plating1} and this elemental lithium
chemically decomposes the SEI there, our work predicts that organic SEI
components, not just additional solvent molecules, may react rapidly there.

Extrapolating from our results, we speculate that Mn(II) and other transition
metal ions diffusing to the anode surface may decompose organic SEI
components via surface reactions on Mn metal surfaces and burn holes in
the outer SEI layers.  This may be one reason transition metal
ions degrade passivating films.\cite{nidiffus,mndiffus}  

If the anode of interest is purely Li metal, our work suggests that different
SEI components can decompose on its surface, releasing O$^{2-}$ and carbon.
Li$_2$O (and LiF, if a fluoride source is present) would be the passivating
films covering the surface.  OC$_2$H$_4$O$^{2-}$ may also be present.  However,
Ref.~\onlinecite{e2} suggests that OC$_2$H$_4$O$^{2-}$ rapidly reacts with
intact solvent molecules to form oligomers that can subsequently yield
Li$_2$CO$_3$ and LEDC.  Since the latter components are not stable on Li metal
surfaces, OC$_2$H$_4$O$^{2-}$ may eventually be destroyed in multistep
reactions.  The eventual thickness of the Li$_2$O oxide layer, and the fate of
the small amount of carbon on the surface, depend on O$^{2-}$ and C$^{q-}$
diffusion rates inside Li metal, and will be the subjects of future studies.

The variations in reaction barriers for Li$_2$CO$_3$ (001) with different
amount of interfacial Li [Figs.~\ref{fig4}d-e; Fig.~\ref{fig8}h (zero
barrier)], and for an isolated CO$_3^{2-}$ (Fig.~\ref{fig4}f), illustrate
that the Li$_2$CO$_3$ bond-breaking rate can strongly depend on the
chemical environment.  We have focused on the initial decomposition process,
and speculate that, as CO$_3^{2-}$ units are removed by reactions, they leave
voids that can be filled by Li atoms, yielding more interstitials-like Li
outcroppings that can accelerate decomposition of interfacial CO$_3^{2-}$.
As Li$_2$O builds up at the interface, such reactions should be impeded.
Note that other SEI components like Li$_2$VDC may still decompose on the
outside of the Li$_2$O layer.\cite{soto}

If the anode is Li$_x$Si, the situation is more complex.  LEDC films of
varying thickness has been shown in this work and elsewhere\cite{soto} to
decompose on {\it both} Li(100) and Li$_x$Si surfaces.  Li$_2$CO$_3$
decomposition on Li$_x$Si is more challenging due to the possible surface
modifications discussed in the previous paragraph.  (See the S.I.
for the effect of Si-atom doping the surface.) If Li$_2$CO$_3$ also decomposes
readily on Li$_x$Si surfaces, which we consider likely due to the fact
that Li$_x$Si surfaces are covered with Li,\cite{fec} O$^{2-}$ released from
SEI components should form Li$_2$O during charging (see experimental evidence
discussed below).  Upon discharge, Li$_x$Si$_y$O$_z$ becomes thermodynamically
favored over Li$_2$O.\cite{lisio} Hence there could be some transfer of oxygen
from the SEI to silicon.  It is unclear whether the possibly small amount of
carbon released from SEI products forms graphite or remains lithium carbide.  
The initial SiO$_2$ covering the Si surface should transform into lithium
silicate phases at low voltages.\cite{lisio}  It is uncertain whether
Li$_x$Si$_y$O$_z$ formation is fully reversible.\cite{trahey} 

X-ray photoemission spectroscopy (XPS), with varying photon
energies,\cite{edstrom1,edstrom2,edstrom3} and time-of-flight secondary ion
spectroscopy,\cite{evolv1} have been used to perform depth profiling of the
SEI.  Li$_2$O has been found in the SEI region immediately next to
Li$_x$Si in silicon anodes, although this has been attributed to the
transformation of SiO$_2$.\cite{edstrom1,edstrom2}  This Li$_2$O slowly
transforms into LiF, indicating that even the deepest-lying SEI layer can
undergo evolution as cycling proceeds.\cite{edstrom1,edstrom2} Since we have
not considered F-containing compounds in our static DFT work, the competition
between LiF and Li$_2$O in the innermost SEI layer will the subject of
future studies.  We also note that Li$_2$O is unstable with respect to
electron beams and moisture.  Regarding Li$_2$CO$_3$ decomposition,
electrochemical reduction of this species has recently been demonstrated under
restricted conditions.\cite{lqchen}  A promising approach to pinpoint SEI
reactions on the immediate surfaces of active electrode materials may be
UHV-based measurements.\cite{robey}  

Our demonstration of LEDC instability on Li metal surfaces is
another reason that the inner SEI films on Li$_x$Si and Li(s) consists only
of inorganic components and the outer region contains organic compounds.
It is possible that organic components may coat on the electrode surface
after first nucleating in the liquid electrolyte region,\cite{far_shore}
and then transform into inorganic components like Li$_2$O.  In light of this
work, it is of interest to re-examine the stability of newly discovered SEI
components.\cite{leifer}

We stress that most calculations in this work are either static or short AIMD
trajectories.  Our simulation conditions can only be related to dynamical
measurements qualitatively.  For example, the charging or ``C''-rate cannot be
quantified.  However, quasi-static overpotential voltage conditions
can be represented
by creating mismatches between the electronic voltage (related to Fermi
levels) and ionic voltage (related to lithium content).\cite{solid}
In anode materials like Si or Sn, SEI thickness and chemical composition
(e.g., organic vs.~inorganic) can change during charge/discharge; the
passivating film can even crack, exposing pristine anode materials, and needs
to be reformed from scratch (i.e., the surfaces are at ``zero SEI thickness'').
DFT-based data will be crucial to parameterize multiscale models to address
these issues.

Finally, in terms of modeling, this work highlights the challenges facing
systematic treatment of buried solid-solid
interfaces.\cite{sodeyama,holzwarth1,santosh,sumita}  Imaging techniques
have yet to achieve the resolution needed to elucidate the atomic details
that can be used as starting points for modeling such interfaces.  Therefore
a suite of surface models ranging from pristine to defect-incorporated
are considered.  Other factors to be considered in the future include
strain induced by inserting/removal Li during charge/discharge.  The
low bulk modulus Li metal and high ``x'' Li$_x$Si materials considered in
this work are not expected to have their electronic properties or reactivities
strongly affected by strain, but amorphous Si may.  Chemical/spatial
heterogeneities, facet dependence, and mechanical deformation also need to
be addressed.  Systematization of interfacial modeling will likely emerge
after more modeling studies of individual sets of interfaces have highlighted
the commonalities and differences.

\section{Conclusions}

\label{conclusions}

In this work, we consider three criteria of SEI stability: thermodynamic,
electrochemical, and interfacial.  Two key SEI components, lithium carbonate
(Li$_2$CO$_3$) and lithium ethylene dicarbonate (LEDC), are predicted to be
electrochemically stable but thermodynamically unstable.  The fundamental
reason for thermodynamic instability is likely the existence of carbon
atoms in high formal charge states under highly electrochemically reductive
conditions.  By this definition, we conjecture that most carbon-containing
SEI components may actually be thermodynamically unstable.

Interfacial reactions can trigger the decomposition of these SEI products.
Our static DFT calculations focus on the interfaces between these SEI
components and two model surfaces.  Li(100) and a-Si are chosen to
represent two extremes of reactive anode.  LEDC proves to be fragile,
decomposing on Li(100) with reaction barriers predicted to be significantly
lower than 0.92~eV.  This suggests that decomposition will occur within
battery operation (i.e., about one-hour) timescales.  LEDC decomposition on
a-Si surfaces is predicted to be slower than the 1-hour threshold time frame.
Li$_2$CO$_3$ decomposition is observed on Li metal surfaces.  The reaction
barrier is below the 0.92~eV threshold in the presence of excess Li atoms at
the interface.  Therefore we predict that a thin layer of Li$_2$O, of at
present unknown thickness, exists on lithium metal surfaces below the
rest of the SEI components, unless Li$_2$O is subsequently converted into
LiF.  Similar Li$_2$CO$_3$ degradation reactions and Li$_2$O formation are
speculated to occur on high Li-content Li$_x$Si surfaces.  DFT-based molecular
dynamics simulations are also used to demonstrate that organic SEI components
spontaneously decomposes on explicit Li$_x$Si surfaces.  Carboxide groups are
found to be kinetically stable in the absence of liquid electrolytes.  Our
predictions have mulitple implications for SEI evolution during cycling on
Li metal and Si anodes, and undesirable Li-plating on the SEI.

\section*{Acknowledgement}

We thank Oleg Borodin, Byron Konstantinos Antonopoulos, Wentao Song, and
Janice Reutt-Robey for discussions and input.  
Sandia National Laboratories is a multiprogram laboratory
managed and operated by Sandia Corporation, a wholly owned subsidiary of
Lockheed Martin Corporation, for the U.S.~Department of Energy's National
Nuclear Security Administration under contract DE-AC04-94AL85000.  
This work was supported by the Assistant Secretary for Energy Efficiency
and Renewable Energy, Office of Vehicle Technologies of the U. S. Department of
Energy under Contract No. DE-AC02-05CH11231, Subcontract No. 7060634 under the
Advanced Batteries Materials Research (BMR) Program.


\begin{references}
\bibitem{note1}
A previous work on crystalline Li$_2$CO$_3$ finds that adding $e^-$
and Li$^+$ interstitual (i.e., a Li atom) constitute a low-energy
defect under battery operating conditions.\cite{qi2}  For the purpose of
estimated reduction potential, we have not considered interstituals.
\bibitem{co2}
The release of ``CO$_2$'' should not be considered oxidation.  CO$_2$ molecules
are electrochemically reduced at $\sim -2$~V vs.~SHE ($\sim 1$~V
vs.~Li$^+$/Li(s)) in organic solvents.\cite{co2_ref}
\bibitem{note2}
We have considered displacing the Li$_2$CO$_3$ and the Li(100) slabs relative
to each other in the lateral directions.  The entire Li$_2$CO$_3$ slab is
moved in the $x$- or $y$-directions by increments of 1~\AA, and the geometry
is optimized.  The resulting energy change per surface Li$_2$CO$_3$ formula
unit is small, in the range $-0.01$~to~$0.034$~eV.
\bibitem{note3}
In constast, in the LEDC simulation cells of Fig.~\ref{fig3},
the C$_2$H$_4$ backbones are exposed to the surface.  Organic solvents will
exhibit far less preferential orientations on this surface and their effects
on the electronic voltage will be less than on Li$_2$CO$_3$ surfaces.
\end{references}
\end{document}